\documentclass[prb,twocolumn,showpacs,floatfix,superscriptaddress]{revtex4}
\usepackage[utf8]{inputenc}
\usepackage{braket}
\usepackage{amsmath}

\usepackage{graphicx}

\newcommand{\eps}{\epsilon}

\newcommand{\dee}[0]{\mathrm{d}}
\newcommand{\idee}[0]{\,\dee}
\newcommand{\gdiff}[3]{\frac{#1 #2}{#1 #3}}
\newcommand{\diff}[2]{\gdiff{\dee}{#1}{#2}}

\newcommand{\colorcaption}[1]{\caption{(Color online) #1}}
\graphicspath{{fig/}{script/}}
\begin{document}
\title{Electronic shell structure and chemisorption on gold
  nanoparticles}
\author{A.\ H.\ Larsen}
\affiliation{Center for Atomic-scale Materials Design, Department of
Physics \\ Technical University of Denmark, DK - 2800 Kgs. Lyngby, Denmark}
\author{J.\ Kleis}
\affiliation{Center for Atomic-scale Materials Design, Department of
Physics \\ Technical University of Denmark, DK - 2800 Kgs. Lyngby, Denmark}
\author{K.\ S.\ Thygesen}
\affiliation{Center for Atomic-scale Materials Design, Department of
Physics \\ Technical University of Denmark, DK - 2800 Kgs. Lyngby, Denmark}
\author{J.\ K.\ Nørskov}
\affiliation{SUNCAT Center for Interface Science and Catalysis, SLAC
National Accelerator Laboratory \\ 2575 Sand Hill Road, Menlo
Park, CA 94025, USA}
\author{K.\ W.\ Jacobsen}
\affiliation{Center for Atomic-scale Materials Design, Department of
Physics \\ Technical University of Denmark, DK - 2800 Kgs. Lyngby, Denmark}

\date{\today}

\pacs{36.40.Cg, 36.40.Mr, 36.40.Jn, 34.35.+a}

\begin{abstract}
We use density functional theory (DFT) to investigate the electronic
structure and chemical properties of gold nanoparticles.  Different
structural families of clusters are compared.  For up to 60 atoms we
optimize structures using DFT-based simulated annealing.  Cluster
geometries are found to distort considerably, creating large band gaps
at the Fermi level.  For up to 200 atoms we consider structures
generated with a simple EMT potential and clusters based on
cuboctahedra and icosahedra.  All types of cluster geometry exhibit
jellium-like electronic shell structure.  We calculate adsorption
energies of several atoms on the cuboctahedral clusters.  Adsorption
energies are found to vary abruptly at magic numbers.  Using a
Newns-Anderson model we find that the effect of magic numbers on
adsorption energy can be understood from the location of
adsorbate-induced states with respect to the cluster Fermi level.
\end{abstract}

\maketitle
\section{Introduction}
A major theme in advanced materials design today is the possibility to
modify and change materials properties through structuring at the
nanoscale. The applications can be as diverse as optimizing the size
of metal nanoparticles to catalyze certain chemical
reactions\cite{Valden:1998gx} or the structuring of surfaces and interfaces for
optimal light absorption in photovoltaic
devices\cite{Atwater:2010kq}. In very broad terms the interesting
possibilities arise when the structures reach a scale comparable to
the wavelengths of the relevant quantum particles (electrons, plasmons
or photons). 

In this work we investigate theoretically the properties of freestanding metal nanoparticles
made of gold in particular.  The purpose is to improve our understanding of the
relationship between cluster size and a range of electronic and chemical
properties. Different aspects of this has been investigated in
numerous studies.  
See for example the review by Baletto and Ferrando.\cite{RevModPhys.77.371}
What is special here is that we
investigate the cluster properties over an---for electronic structure
calculations---unusually large size range and for many
different cluster structures. The hope is thereby to get a more
complete picture of the general trends in the cluster behavior.

For transition metals with partially filled d-bands, cohesive energies will be
dominated by the effect of the d-states\cite{pettifor1978}.
Because of the short range of the d-states, their contribution to the
cluster energy is determined mostly by the local arrangement of neighboring
atoms.  Facet types and local atomic packing can therefore be expected to be
particularly important factors in the structures of transition metal clusters 
with partially filled d-bands.

The effect of the partially occupied d-band disappears for noble
metals and alkali metals.  Instead the long-range s-electrons, which
hybridize in a more complex manner, yield the primary contribution to
the cluster energy.  The optimal structure will not be
determined by optimizing the local structure around each atom, but
rather by optimizing the global geometric structure to obtain the most
desirable electronic structure of the delocalized electron cloud.  The
result is a much more complicated interplay between electronic and
geometric structure.

Small free-standing gold clusters have been theoretically shown to
possess very diverse ground-state geometries depending on cluster
size.  Examples are planar, cage-like and tube-like
structures\cite{haekkinen_electronic_2003_1,%
  PhysRevB.70.165403,PhysRevB.74.165423,Zhao20101033}.

The s-electron hybridization can be interpreted in terms of a jellium
model which regards the whole cluster as a spherical superatom.  The s
electrons organize into global shells, resulting in electronic ``magic
numbers'' when shells are filled.  Magic numbers at 2, 8, 18, 20, 34, 40, 58,
\ldots, have been observed as particularly stable alkali metal
clusters\cite{knight-clemenger-sodium1984,PhysRevA.65.063201}
with large band gaps in agreement with theory.  For alkali metals,
magic numbers attributed to both electronic and geometric shell
structures have been observed for clusters with thousands of
atoms, with geometric shells dominating
beyond 2000 atoms\cite{martinshellsofatoms}.
Larger Au clusters are believed to
form icosahedra, decahedra or truncated octahedra depending on size and
temperature\cite{barnard_equilibrium_2005,barnard_nanogold:_2009,%
  doye_entropic_2001}.

In this work we consider several series of clusters based on different
generation procedures and structural motifs.  We calculate structures
of smaller clusters using simulated annealing with density functional
theory (DFT) and for larger clusters using effective medium
theory\cite{PhysRevB.35.7423,Jacobsen1996394} (EMT).  Using DFT we
compare the energy and electronic structure of optimized clusters
with the commonly considered regular icosahedral and cuboctahedral
structures.  For the cuboctahedra we identify trends in reactivity by
considering adsorption of different atoms.  The geometric similarity
of clusters based on cuboctahedra and icosahedra allows us to isolate and 
study size-dependent
effects on chemistry.  The price of this simplification is that
individual calculations do not represent globally optimal structures.
Hence we focus on trends that are general enough to be significant
outside the model systems.

\section{Computational methods}
All electronic structure calculations are performed with the
real-space DFT code GPAW\cite{mortensen_real-space_2005,enkovaara_electronic_2010} using the
RPBE\cite{RPBE} functional for exchange and correlation.  GPAW uses
the projector augmented wave (PAW) method\cite{blochl1994}, and offers
an accurate real-space representation of the Kohn-Sham orbitals along
with an efficient basis set of localized atomic orbitals\cite{larsen_localized_2009}.
The calculations presented here are performed with the atomic orbitals
using a double-zeta polarized (DZP) basis set.  All calculations on
clusters are spin-paired and use the scalar-relativistic atomic PAW
setups and basis sets supplied with GPAW.
The Au setup contains 11 valence electrons.

In our calculations the cluster is centered in a non-periodic
orthorhombic cell with 5.0\,Å vacuum along each axis.  We use a
grid-spacing of 0.2\,Å.

We do not apply any basis set superposition error correction, so the
values of adsorption energies are not necessarily accurate.  However
in comparing the bonding of an adsorbate to clusters of different
sizes, the local structure around the adsorbate highly similar for all
clusters, and the basis set error is consequently roughly the same for
all clusters.  Therefore variations in adsorption energies are subject
to a much smaller error.

Pulay density mixing\cite{pulay} is used to speed up convergence of
the self-consistency loop.  Electron occupations are smeared by a
small Fermi temperature of 0.01\,eV, which helps speed up convergence.

Structure optimizations are performed using the BFGS algorithm as
implemented in the Atomic Simulation Environment\cite{ase} (ASE), and
terminate when no force on any atom is larger than 0.075\,eV/Å.
\section{Cluster geometry}
Systematic calculation of lowest-energy structures from first principles 
is computationally
very expensive. Previous studies of structures and properties of Au
clusters have therefore usually been limited to a few dozens of
atoms.\cite{Zhao20101033,xiao_structural_2006,bokwon-sizedependent,snow_size-induced_1998}
Here we focus mainly on larger clusters
which are quite challenging to systematically optimize and
characterize, but which are clearly of interest both from a conceptual
point of view and in applications like catalysis.  The transition from
smaller clusters over larger clusters to bulk-like behavior has been
studied recently by Kleis \textit{et al.}\cite{kleis-quantumsize} and
the results presented here can be seen as a supplement and expansion
of this study.

In this work we compare clusters generated by several different
procedures.  For the smallest clusters we perform simulated annealings
using DFT to obtain realistic structures. This is clearly the most
realistic and theoretically satisfactory method since the same energy
landscape is used to define the cluster geometry or shape as is used
to subsequently study the bonding of the cluster and the chemical
properties. However, for reasons of computer time this approach cannot
be generally applied for larger clusters.

Larger clusters are studied using DFT, but with the structures being
determined by simulated annealing with a classical EMT potential.  As
this EMT potential does not incorporate explicit electronic structure,
the structures generated by this method will have no information about
potential electronic shell effects but only of atomic shell effects
related to atomic packing of the clusters.

We finally construct clusters based on prescribed cuboctahedral and
icosahedral shapes.  The simplicity of the fcc-based cuboctahedral
structures allows us to study adsorption of atoms in a way which
preserves the local geometry around the adsorbate for different
cluster sizes.  This allows us to separate the effect of local
geometry from that of the electronic structure of the cluster, which
would not generally be possible if the cluster were based on a global
minimum search.  The comparison of distinct types of structures will
help determine how properties of clusters depend on structure versus
size.

\subsection{Simulated annealing with DFT}
For the smallest clusters ($N$=6--60) we calculate realistic geometries
using DFT with coarse parameters.

For each size of cluster we perform a rough simulated annealing based
on molecular dynamics (MD) to find the optimal structure.  We use a
Langevin thermostat to regulate the temperature from 750\,K to 300\,K.
For a cluster of size $N$ we lower the temperature by 1\,K for every $5
+ N / 2$ timesteps of length 24\,fs.  The timestep is too large to have
accurate energy conservation during the optimization.  This can cause
unrealistic behavior when the atoms move quickly, but is not likely to
affect the results of an annealing where the result is mostly
determined at lower temperatures.  The optimization is performed with
a very coarse grid spacing of 0.24\,Å.

At the end of the MD simulation we perform a structure optimization
with normal DFT parameters using the BFGS algorithm such that the
structure is guaranteed to be a local minimum.

\begin{figure}
  \centering
  \includegraphics[width=0.17\textwidth]{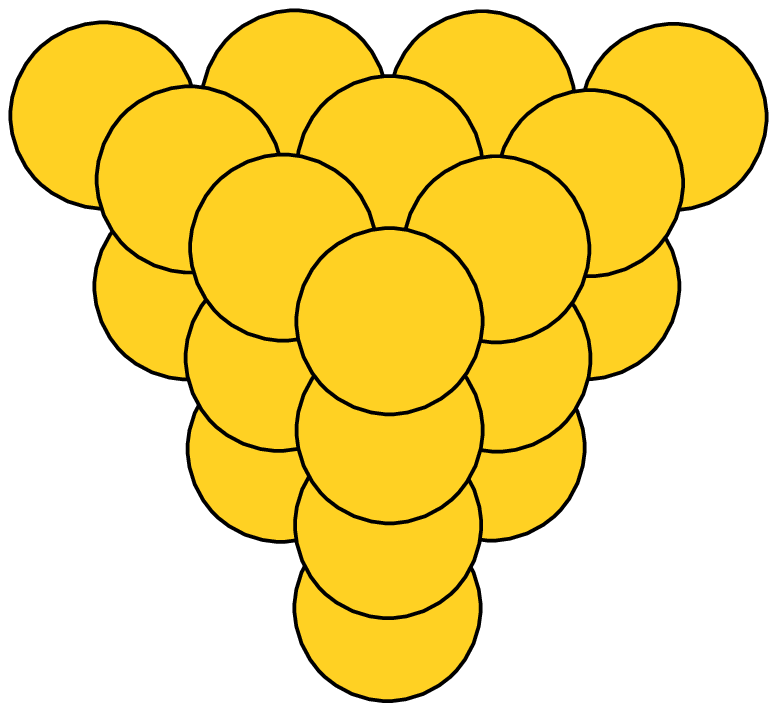}
  \includegraphics[width=0.17\textwidth]{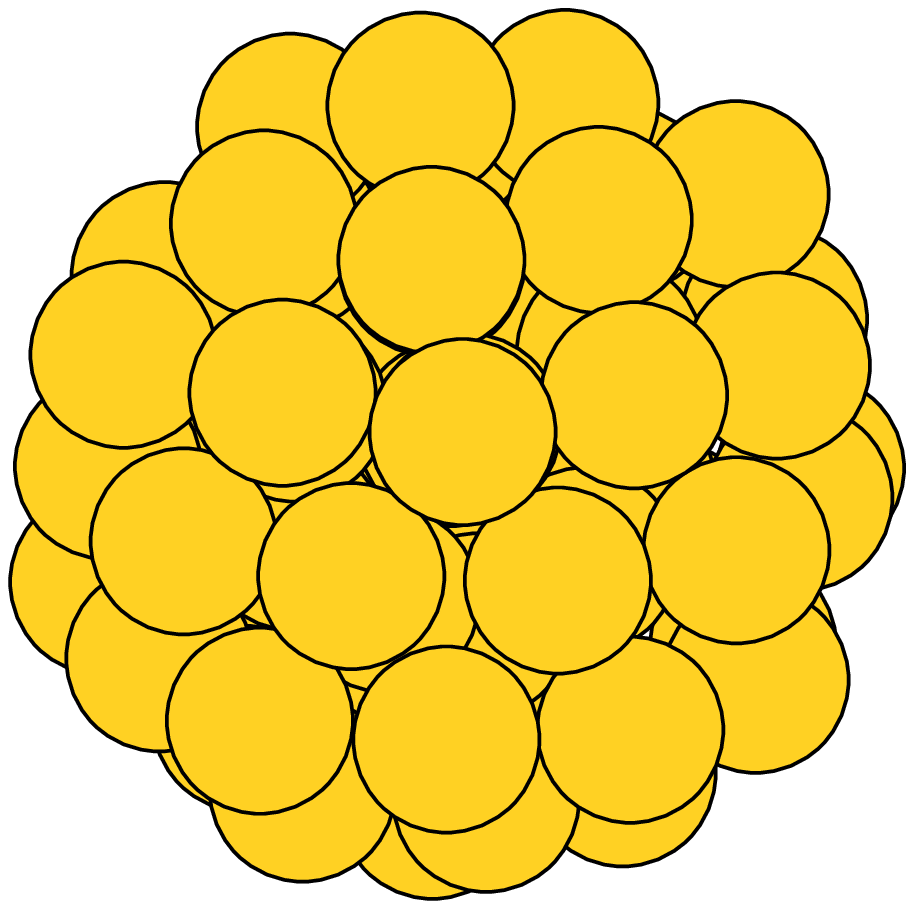}
  \colorcaption{20 (left) and 58-atom (right) clusters obtained by
    simulated annealing with DFT.}
  \label{fig:dft-annealed-geometry}
\end{figure}

These optimizations produce planar and tetrahedral structures
in qualitative agreement with previous findings\cite{Ferrighi:2009hy,%
  PhysRevB.73.235433, PhysRevB.70.165403, Xiao:2004gj}, while larger 
structures
tend to be irregular but with some well-formed facets.  Due to
the short annealing times, the larger structures are unlikely to be global 
optima.  Figure~\ref{fig:dft-annealed-geometry} shows the 20-atom
tetrahedron and the 58-atom cluster obtained with this method.  Notice
on the 58-atom cluster the imperfect five-fold symmetry center
reminiscent of those found on icosahedral clusters.

\subsection{Simulated annealing with EMT}
For a larger range of clusters ($N$=6--200) we calculate structures
using a simple EMT potential\cite{PhysRevB.35.7423,Jacobsen1996394}
implemented in ASAP\cite{asap}.  This potential is designed to provide
reasonable descriptions of elastic and cohesive properties.  It is a
classical potential and as such contains no explicit description of
electronic behavior.

For each size of cluster we perform a simulated annealing wherein the
temperature varies from above the melting point (1337\,K for Au) to
200\,K with 600 MD steps of 6.0\,fs for each 1\,K decrease in
temperature.  The resulting structures frequently have 5-fold symmetry
centers surrounded by 111 facets, resembling partially formed
icosahedra or decahedra.  Again, at the end of the simulated annealing
we perform a BFGS structure optimization with the usual DFT
parameters.

\subsection{Atomic shell structures} \label{sec:geometry-layer-based}

\begin{figure}
  \begin{minipage}{0.22\linewidth}
  \includegraphics[scale=0.17]{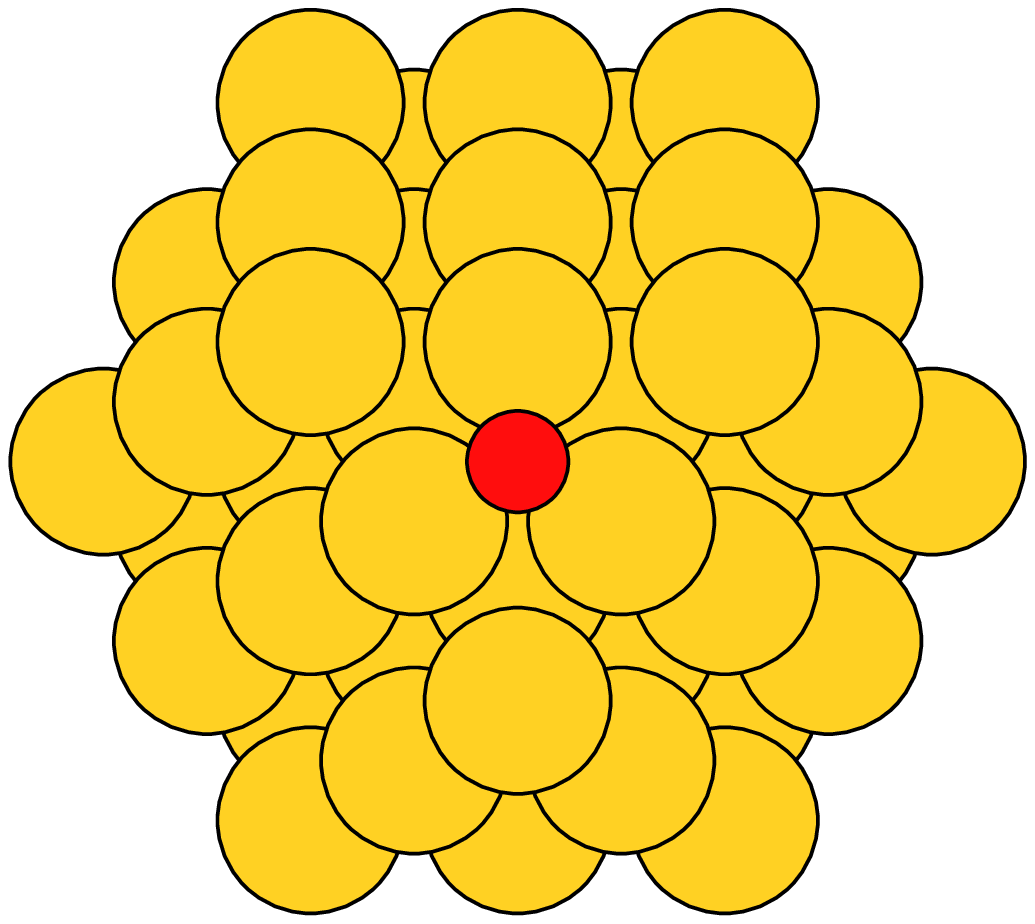}
  \end{minipage}
  \begin{minipage}{0.33\linewidth}
  \includegraphics[scale=0.17]{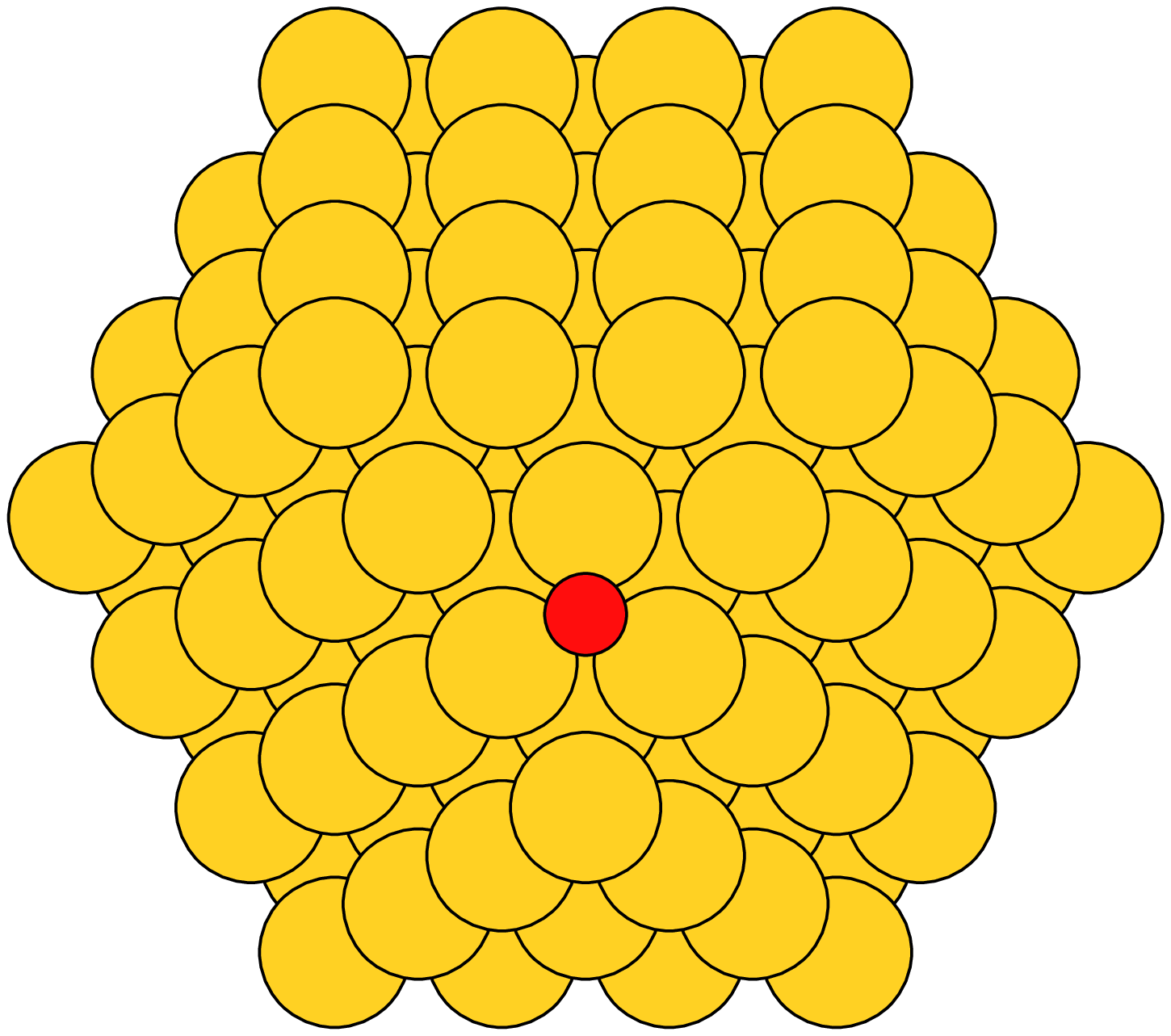}
  \end{minipage}
  \begin{minipage}{0.40\linewidth}
  \includegraphics[scale=0.17]{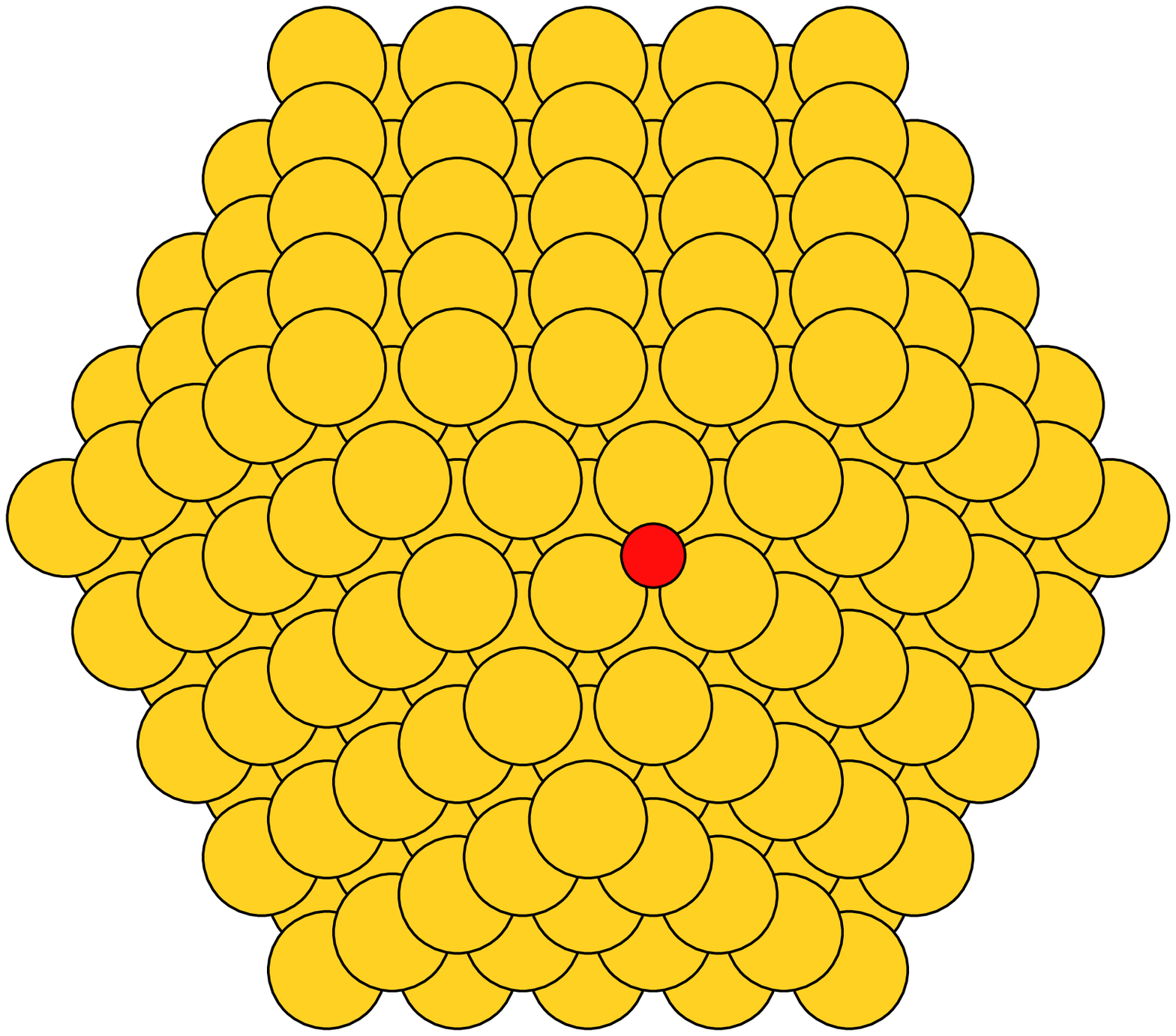}
  \end{minipage}\\
  \begin{minipage}{0.22\linewidth}
  \includegraphics[scale=0.17]{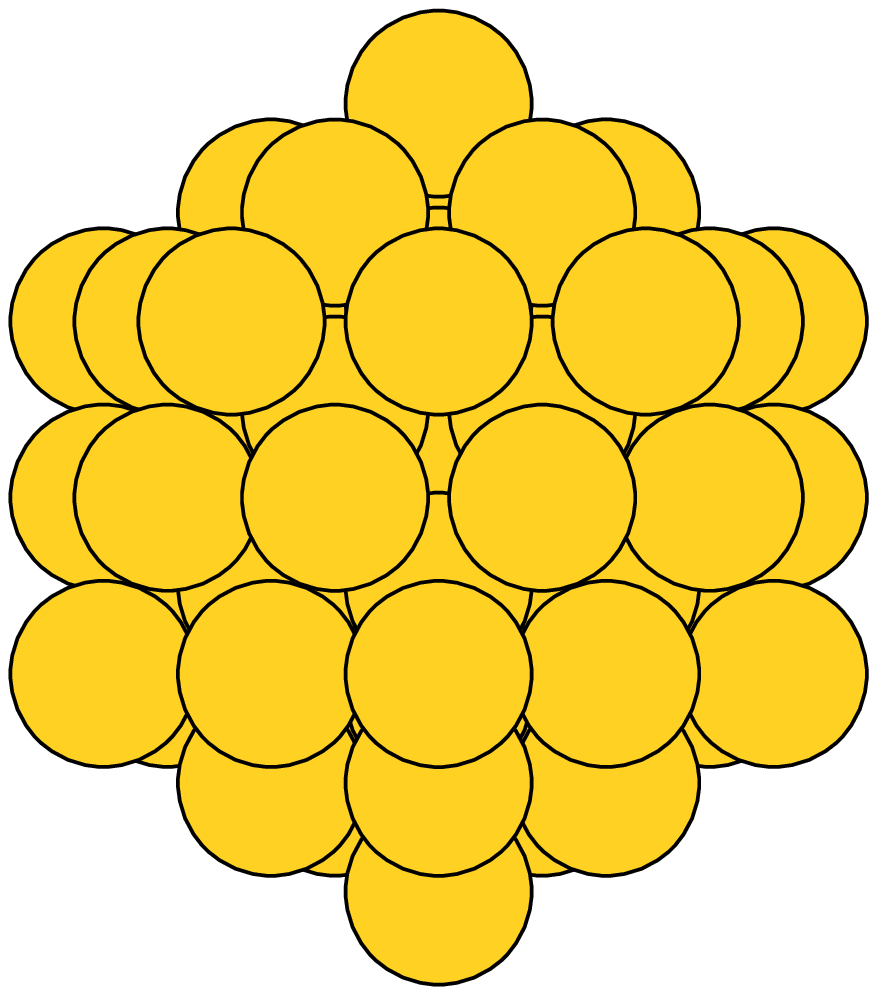}
  \end{minipage}
  \begin{minipage}{0.33\linewidth}
  \includegraphics[scale=0.17]{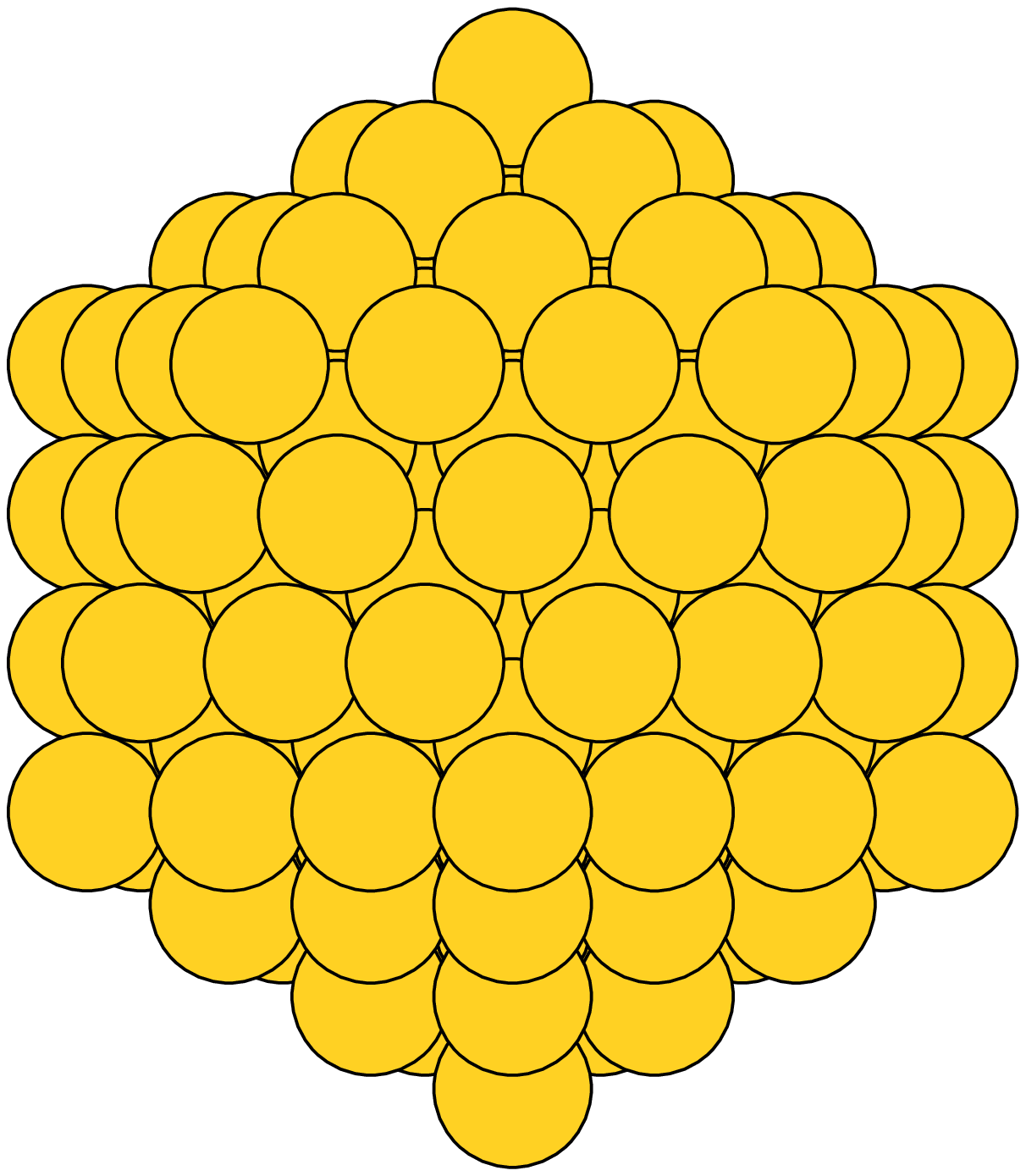}
  \end{minipage}
  \begin{minipage}{0.40\linewidth}
  \includegraphics[scale=0.17]{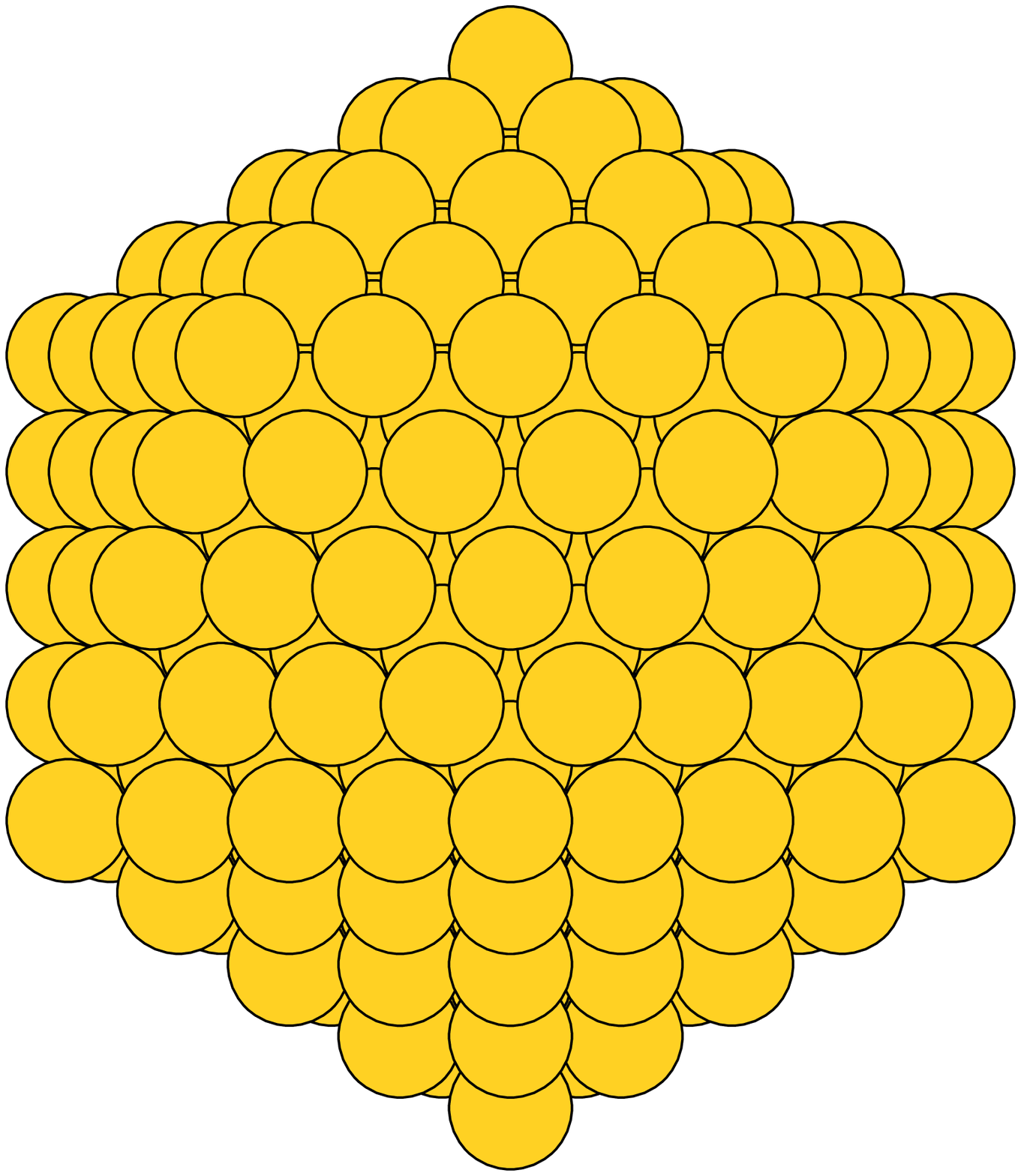}
  \end{minipage}
  \colorcaption{Cuboctahedra (top) and icosahedra (bottom) with 55, 147 and
    309 atoms.  For the cuboctahedra, an O atom is shown at the fcc
    site closest to the center of an 111 facet.}
\end{figure}

Finally we consider the cuboctahedral and icosahedral series of
structures.  Each cluster can be constructed geometrically from the
previous one by adding one complete shell of atoms.  The cuboctahedra
and icosahedra have closed atomic shells at the same numbers.  The
first few geometric shell closings are $N$=13, 55, 147, 309 and 561.
\begin{figure}
  \centering
  \includegraphics[width=0.5\textwidth]{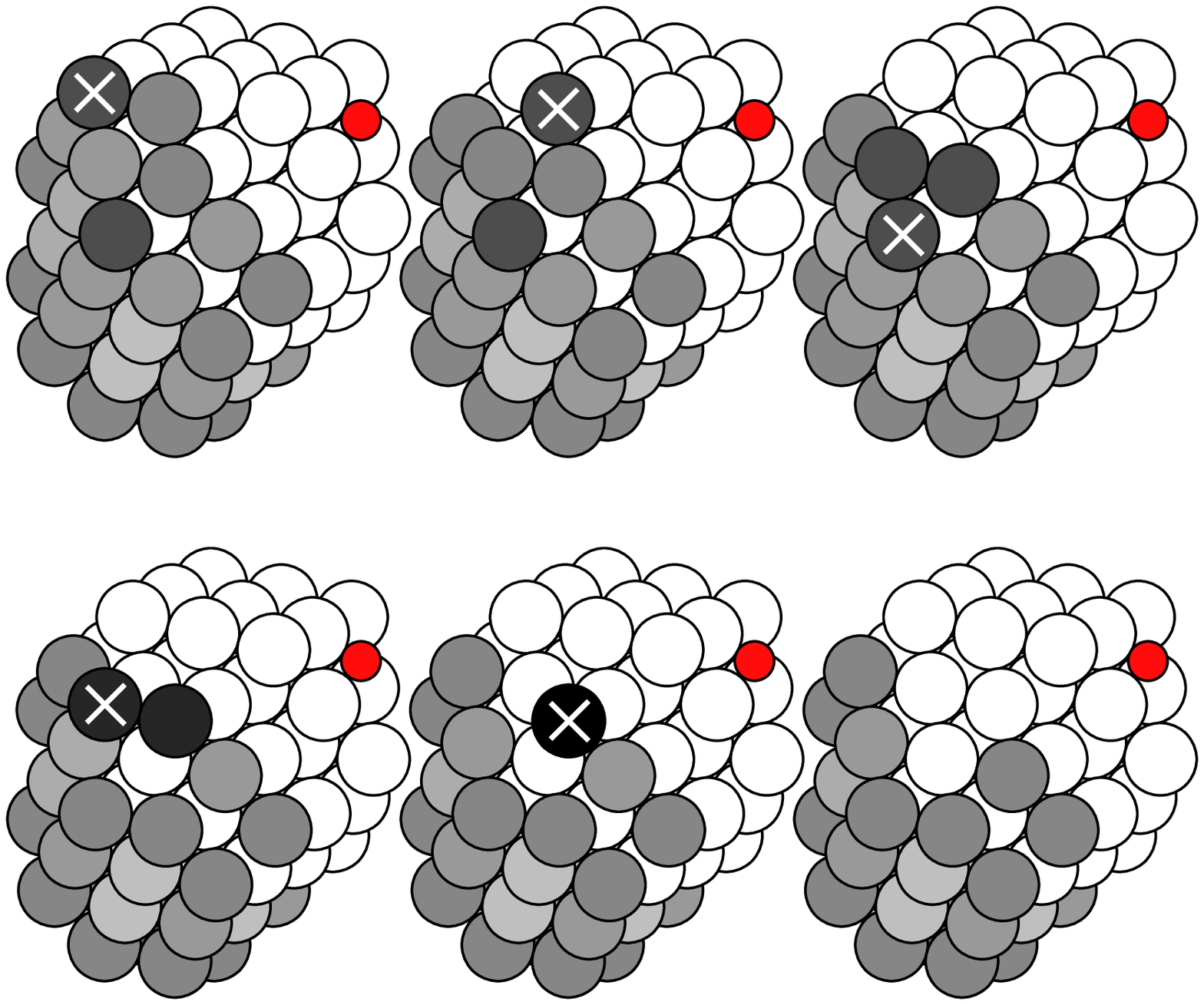}
  \colorcaption{Generation of regular clusters with different numbers
    of atoms.  The white atoms belong to the 55-atom cuboctahedron,
    while the grey atoms are stripped off one by one as marked with a
    cross.  Removable atoms with lower coordination numbers have
    darker shades of grey, and at each step one of the
    lowest-coordinated atoms is removed at random.  An oxygen atom is
    shown at the adsorption site.}
  \label{fig:peel}
\end{figure}

We would like to study the chemical properties of clusters by
calculating adsorption energies of atoms on clusters of
different sizes.  A systematic comparison can be made if we ensure that
the local geometry around the adsorbate remains identical for all
clusters independent of size.

For the cuboctahedra we generate clusters with different numbers of
atoms by stripping off atoms one by one from one cuboctahedron until
only the next smaller cuboctahedron remains.  For each step, the next
atom to be removed is chosen at random amongst those that have the
lowest coordination numbers and are not part of the smaller
cuboctahedron.  This procedure is shown on Figure \ref{fig:peel}.

As mentioned, to obtain adsorption energies that can be meaningfully
compared across the different clusters, we must avoid changing the
immediate environment of the adsorbate when removing atoms.  For this
reason we do not simply remove the outermost shell.  Instead we choose
an adsorption site on a particular facet, then remove atoms as
necessary on the other sides of the cluster such that the local facet
is changed only minimally.

This method preferentially strips off corner atoms and atoms on the
most open facet, opening a new facet only when necessary.  This avoids
very unphysical geometries.  We run the calculations multiple times
using a pseudorandom number generator with different random seeds,
yielding a small ensemble that shows the dependence of cluster
properties on the randomization.

A similar procedure can be applied to the icosahedra.  However in the
icosahedra, the distance between atoms in successive layers is
different from the distance between atoms within the same layer, which
means the local geometry around an adsorbate cannot always be
preserved as for the cuboctahedra.
\section{Intrinsic properties of clusters}

In the following we study energies and electronic properties of the
different types of clusters.  Note that all clusters have been relaxed
using DFT such that all structures are local minima corresponding to
the same force method, and have directly comparable total energies.

\subsection{Cluster structure and stability}

\begin{figure}
  \centering
  \includegraphics{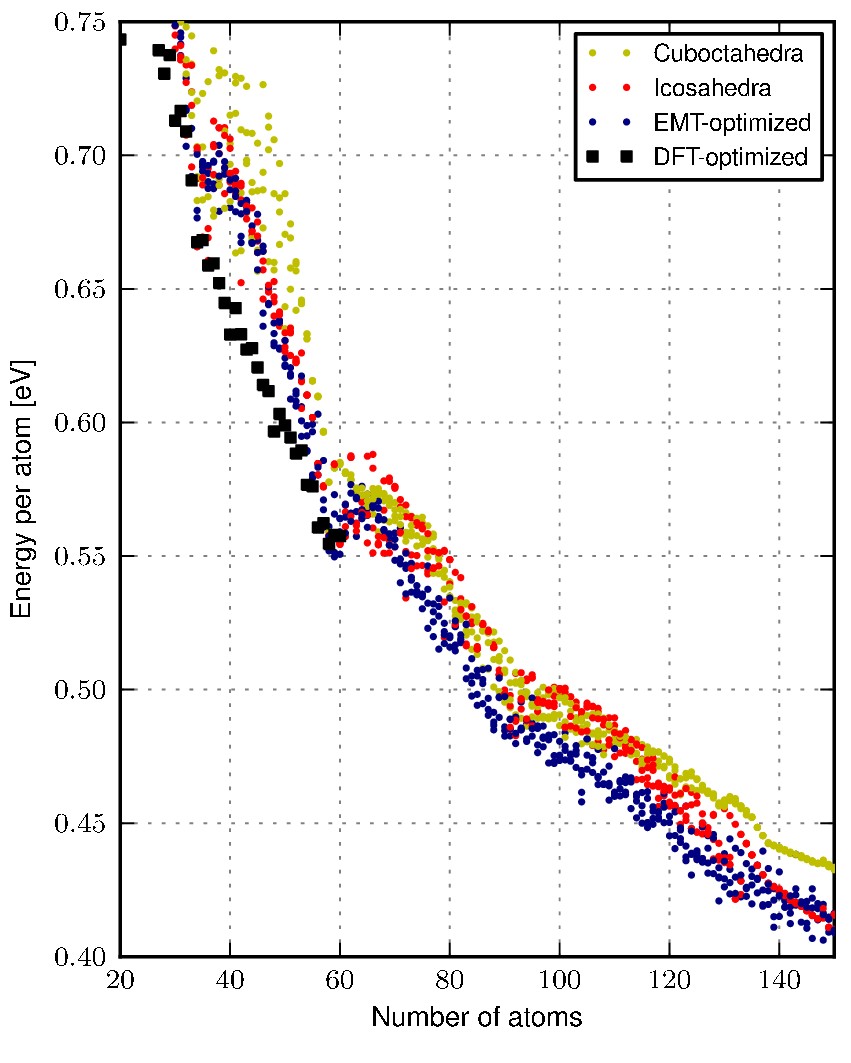}
  \colorcaption{Energy per atom for different Au cluster types as a
    function of cluster size.  The energy reference corresponds to
    bulk Au.}
  \label{fig:energyperatom}
\end{figure}

Figure \ref{fig:energyperatom} compares the energy per atom for Au
clusters of the different types.  EMT-optimized and regular clusters
have been generated multiple times from different pseudorandom number
sequences, yielding four data\-points for each cluster size.  The
DFT-optimized structures generally have energies lower than or equal
to the other methods followed by EMT.  Among the regular
structures, icosahedra usually have lower energies than cuboctahedra.
Even where the regular clusters have closed geometric shells
($N$=55 and 147), they are less stable than the structures obtained by
simulated annealing with EMT.

\begin{figure}
  \centering
  \includegraphics{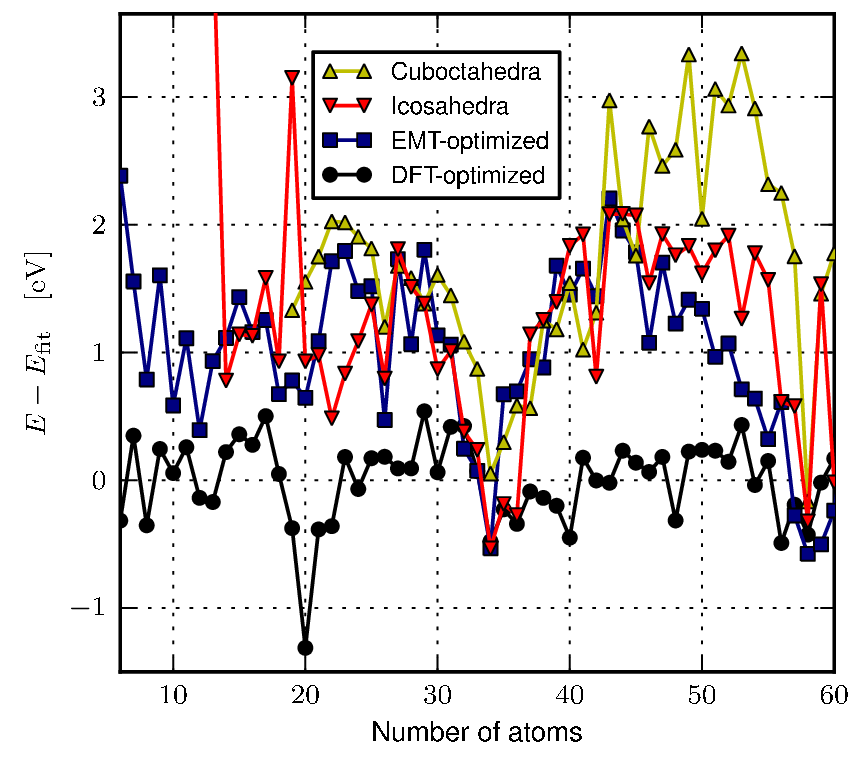}
  \colorcaption{Energy for different Au cluster types as a function of
    cluster size $N$.  A smooth function of $N$
    cf.\ Eq.\ \ref{eq:energyfit} has been subtracted from all
    datapoints for legibility.}
  \label{fig:energyfit.all}
\end{figure}

Prominent kinks in the energy are visible around $N$=34, 58 and 92
atoms.  These are ``magic numbers'' corresponding to major electronic
shell closings.  They are well known in the jellium cluster models of
simple metals\cite{de_heer_electronic_1987, brack_physics_1993,%
  genzken_temperature_1991}, and have also been observed in mass
spectra of noble metal clusters\cite{katakuse_mass_1985}.  The kinks
in energy due to electronic shell structure are robust enough to be
visible for all types of clusters considered.
Figure~\ref{fig:energyfit.all} provides a closer view of the energies
of smaller clusters.  To improve legibility, a smooth function of $N$
of the form
\begin{align}
  E_{\mathrm{fit}}(N) = a_0 + a_1 N^{1/3} + a_2 N^{2/3} + a_3 N,
  \label{eq:energyfit}
\end{align}
is subtracted from all energies.  The coefficients $(a_0, a_1, a_2,
a_3)$ are obtained by fitting the energies of the DFT-optimized
clusters.  For the other kinds of clusters, only the lowest-energy
data\-point found among four attempts is shown for each $N$.  

The DFT-optimized clusters up to 13 atoms are planar except for $N$=10
and 11.  The predicted transition between planar and three-dimensional
structures depends strongly on the approximation of exchange and
correlation, and has been studied more systematically by several
authors\cite{PhysRevB.73.235433,Ferrighi:2009hy}.

The particularly visible feature at $N$=20 is the well-known
tetrahedron\cite{li_au20:_2003}.  Aside from this, particularly stable
clusters are $N$=34, 40, 48 and 56/58.  The EMT-based and regular
structures tend to obtain comparable energies only around the major
magic numbers $N$=34 and 58.  In between the magic numbers, the
EMT-based structures have higher energies than the DFT-optimized
ones by typically 1--2\,eV.

\begin{figure}
  \centering
  \includegraphics{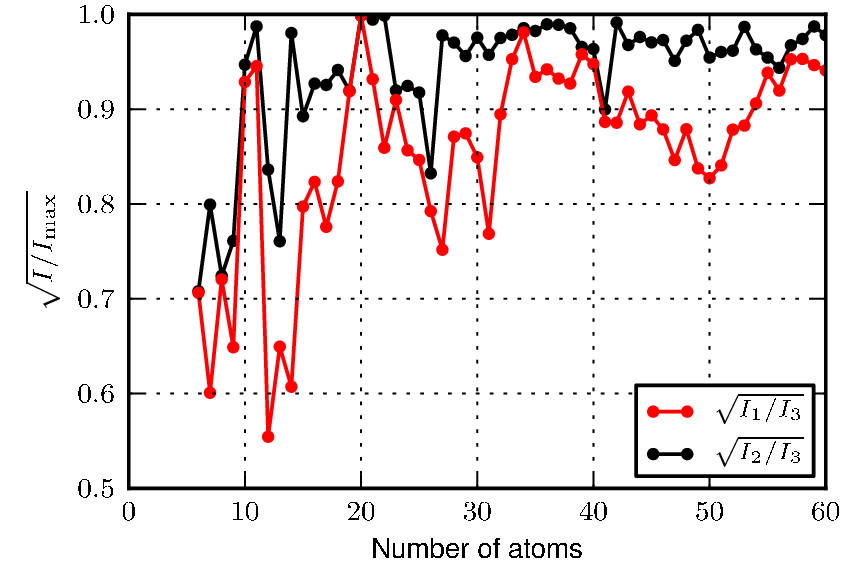}
  \colorcaption{Square roots of ratios $I_1 / I_3$ and $I_2 / I_3$ of the
    three principal moments of inertia $I_1 \leq I_2 \leq I_3$ as a
    function of cluster size, showing deformations of the structures.}
  \label{fig:moments}
\end{figure}

The main structural difference between the DFT-optimized clusters and
other types is that the DFT-optimized clusters systematically deviate
from spherical shapes when doing so is favorable to the electronic
structure.  A rough measure of how spherical a cluster is can be
obtained by considering the moments of inertia.  For each cluster the
three principal moments of inertia $I_1 \leq I_2 \leq I_3$ are
calculated.  Figure~\ref{fig:moments} shows the ratio $\sqrt{I_1 /
  I_3}$ and $\sqrt{I_2 / I_3}$, i.e.\ the square root of the inverse
ratio between the largest principal moment and each of the smaller
ones, as a function of cluster size.  The most symmetric clusters are
found around the magic numbers $N$=20, 34, 40 and 58, 
while intermediate clusters are
deformed considerably.  A similar variation has been predicted for Cu
clusters \cite{obc-kwj-copperclusters,christensen_coupling_1993} 
and in several 
deformable-background jellium models\cite{penzar_electronic_1990_utf,%
  % penzar_self-consistent_1991,
  %yannouleas_electronic_1995,%citations_missing_here
  koskinen_electron-gas_1995}.  Clusters just above magic
numbers are from jellium models expected to be prolate while clusters
below magic numbers are expected to be oblate.  Such a trend is not
clearly visible from our results.  This is most likely due to the
roughness of the optimization procedure combined with the presence of
a physical atomic lattice, modifying the simple model picture.

\begin{figure}
  \centering
  \includegraphics{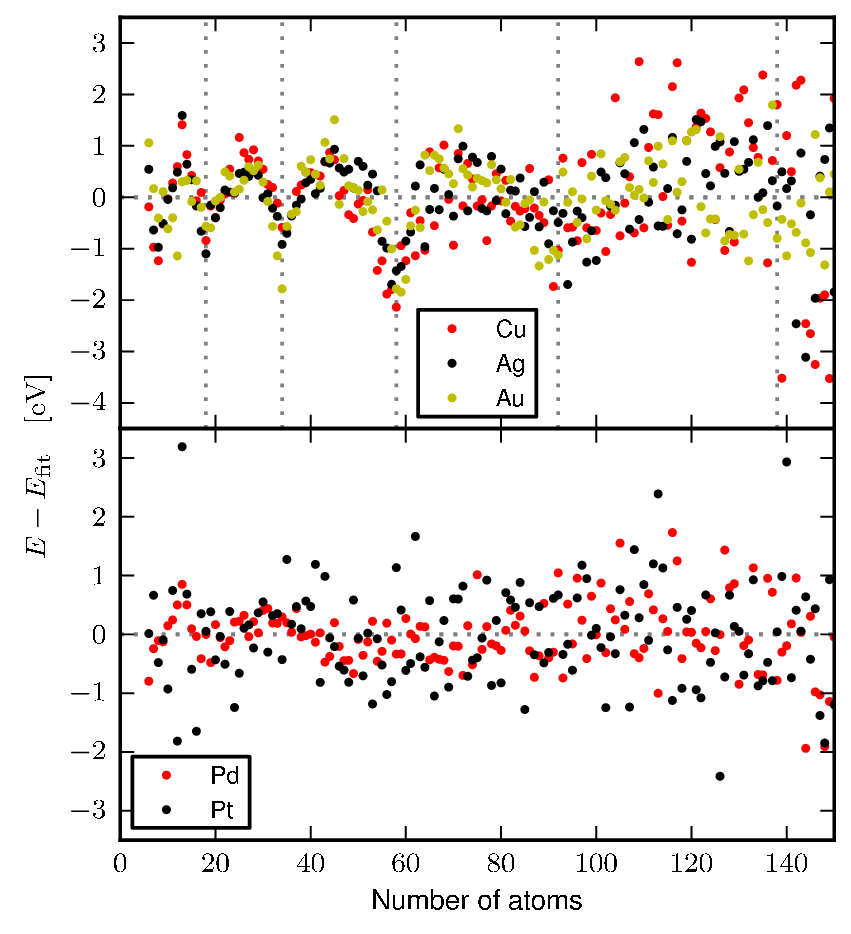}
  \colorcaption{Energy minus fitted trend line for noble metal clusters
    (top) and d-band metals (bottom).  Electronic magic numbers at
    $N$=18, 34, 58, 92 and 138 are indicated.}
  \label{fig:stability-various-species}
\end{figure}

Figure~\ref{fig:stability-various-species} compares the stabilities of
Cu, Ag, and Au (top).  For comparison two other transition metals Pd and
Pt are also shown (bottom).  For each species, the energy is
calculated using simulated annealing with EMT followed by a local
geometry optimization with DFT for $N$=6--200.  A smooth function is
then subtracted by fitting the energies for each element according to
Eq.~\ref{eq:energyfit}, such that the figure shows the deviation from
a smooth trend line.

As for Au, the other noble metal clusters are particularly stable
close to the magic numbers 18, 34, 58 and 92, and to a lesser extent
at 138.  Deviations from the trend line oscillate with a peak-to-peak
variation of around 3\,eV.  Beyond 138, the periodic trend is
gradually obscured by fluctuations which are probably caused by
imperfections in the optimization procedure.

The three noble metals exhibit roughly identical behaviour except in
the region $N < 18$, where Au differs noticeably from Cu and Ag.  In
this case the Au clusters deform considerably from the geometries
found by EMT, tending towards flat structures.  Cu and Ag clusters
remain round.  The tendency of small Au clusters to form planar
structures has been well documented and has been attributed to
relativistic effects causing a contraction of the s-orbitals compared
to the d-orbitals.\cite{krause_ab_2005,pyykko_relativity_1979,%
PhysRevLett.89.033401}
The differences in behavior between noble metals here
appear to be caused exclusively by the relativistic behavior
of Au, and not e.g.\ the location of the d-band in relation to the Fermi
level, which would set Cu and Ag apart.

For Pd and Pt, no magic numbers are observed.  Deviations from the
trend line instead appear to depend on how well-formed the clusters
are, i.e.\ the type and regularity of their facets as seen from visual
inspection of the cluster structures.  Thus the stability of
non-noble transition metal clusters is determined mostly by local
atomic arrangement, corresponding to interactions between the
short-ranged d-electrons.

\subsection{Electronic structure}

\begin{figure}
  \centering
  \includegraphics{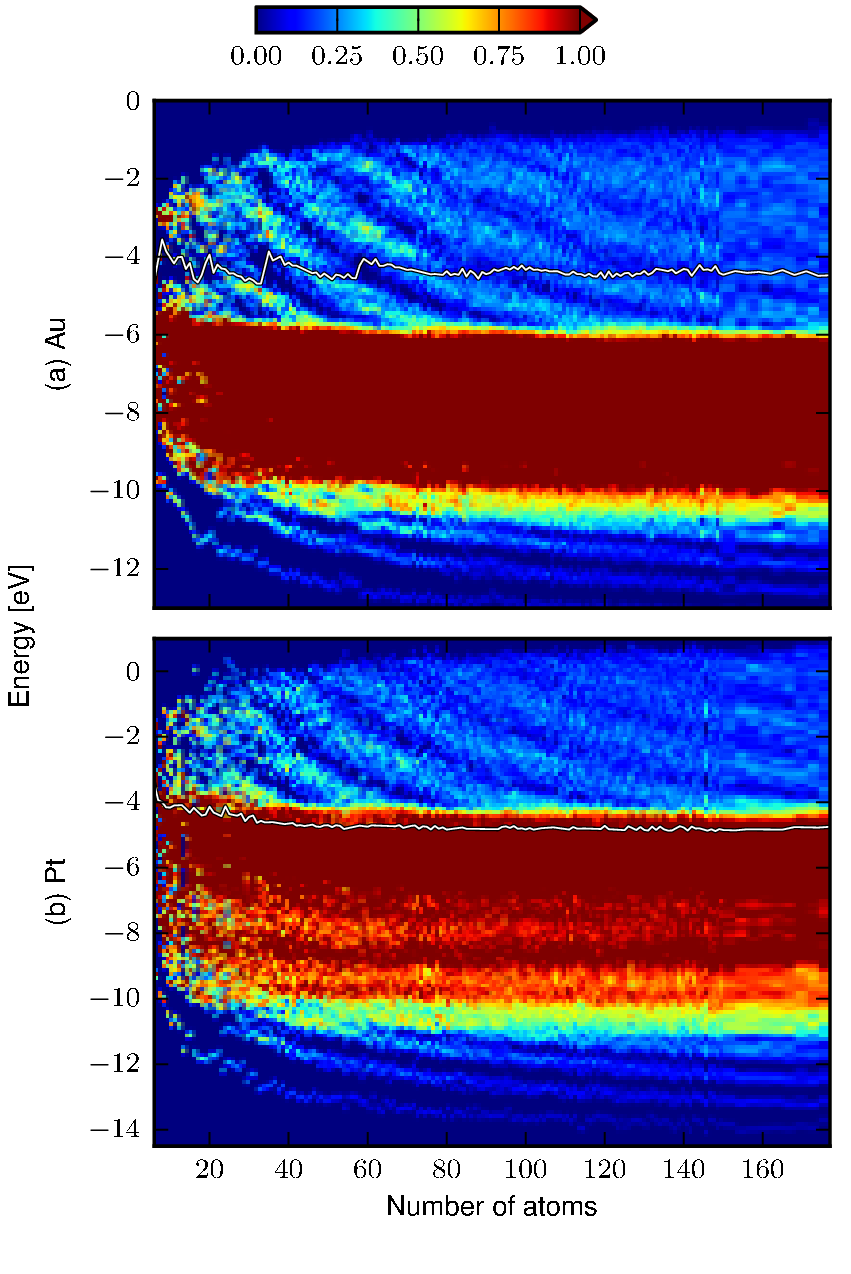}
  \colorcaption{DOS in arbitrary units of EMT-optimized Au clusters (top) 
    and Pt clusters
    (bottom) as a function of cluster size and energy. The line
    indicates the Fermi level.
    Values larger than 1.0 are truncated to 1.0.}
\end{figure}

\begin{figure}
  \centering
  \includegraphics{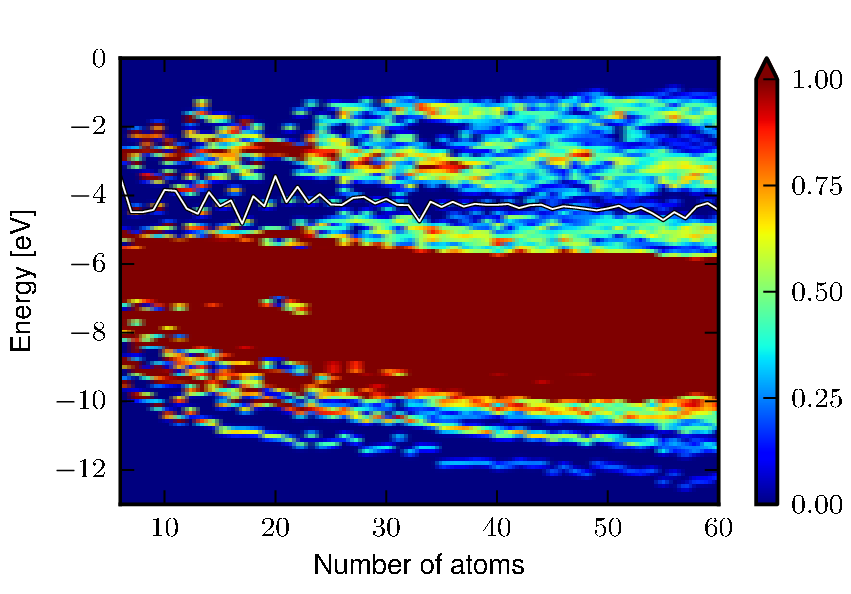}
  \colorcaption{DOS in arbitrary units of DFT-optimized clusters. 
    The line indicates the Fermi
    level.  DOS values larger than 1.0
    are truncated to 1.0.}
  \label{fig:dft-opt-dos}
\end{figure}

Figure \ref{fig:dft-opt-dos} compares the density of states (DOS) per
atom of Au (top) and Pt (bottom) clusters optimized with EMT as a
2-variable function of cluster size and energy.  For each cluster, the
DOS is approximated as a sum of Gaussians of width 0.07\,eV centered on
each energy eigenvalue.

For both Au and Pt the d-states very quickly form the usual
continuous, narrow band which beyond $N$=20 changes only very little.
The s-states split up into multiple electronic shells which are
separated by gaps as in the jellium shell model.  As $N$ increases the
shells gradually broaden to form a continuous band.  Oscillations in
the DOS originating from the shell structure are still clearly visible
even for the largest clusters.

For Au, where the Fermi level is located well above the d-band, the
electronic shells due to the s-electrons are filled one by one as
cluster size increases.  When one shell is full, the Fermi level jumps
to the next higher shell, causing the abrupt shifts in Fermi level and
large band gaps.

The Fermi level for Pt is lodged at the top of the d-band where the
DOS is extremely high.  Therefore no gaps or jumps in the Fermi level
are possible, and the cluster will not exhibit any electronic magic
numbers even though the s-electrons form shells exactly like Au.

Comparing to the DOS of the DFT-optimized clusters on Fig.\
\ref{fig:dft-opt-dos}, the DFT-based optimization consistently creates
very large gaps around the Fermi level for all small clusters.  For
clusters with an odd number of electrons (where the gap is zero
because of spin-degeneracy), a single singly-occupied state is located
at the middle of a symmetric gap.  Similar behavior has been reported for 
Cu clusters\cite{obc-kwj-copperclusters}.
A significant difference compared to the roughly
spherical clusters obtained with EMT is that the shell structure
cannot easily be seen as distinct shells that move down in energy as
the cluster size increases.  Rather there is an accumulation of states
some way above as well as below the Fermi level.  Only close to the
shell closings does the DOS resemble that of the EMT-optimized
clusters.
\begin{figure}
  \centering
  % actually the graph shows stuff for DFT clusters in spite of the filename
  \includegraphics{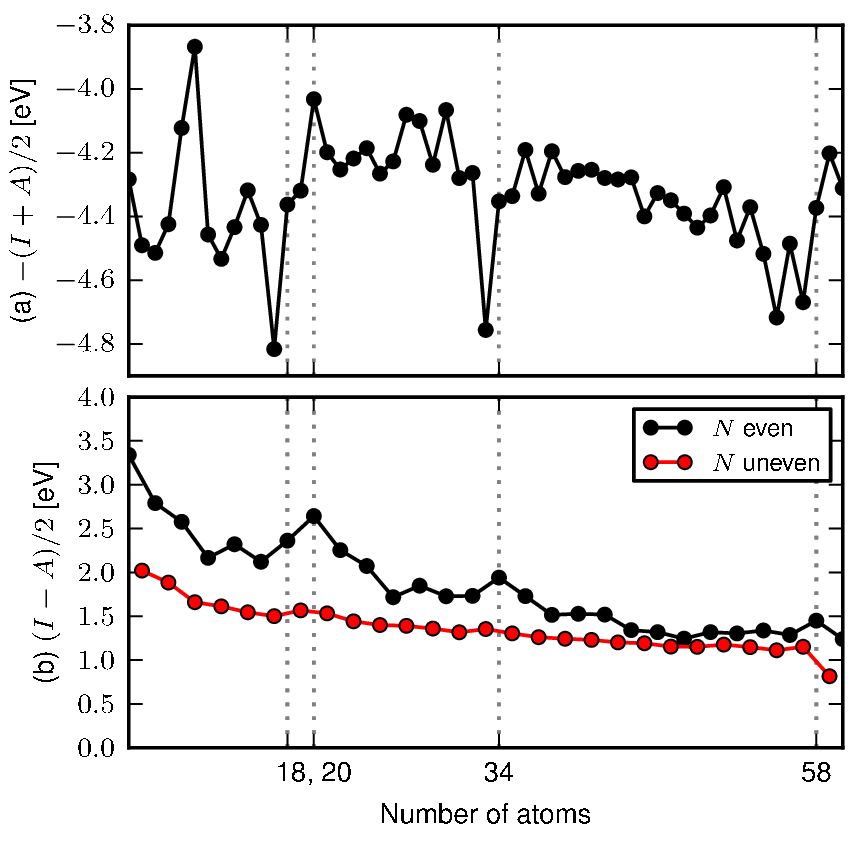}
  \colorcaption{Fermi level (top) and chemical hardness (bottom) calculated
    from ionization potential $I$ and electron affinity $A$ for
    DFT-optimized Au clusters as a function of cluster size.}
  \label{fig:electronproperties}
\end{figure}
A consequence of this is that the abrupt jumps in Fermi level $\eps_F$ seen for
EMT-optimized clusters are less visible for the DFT-optimized ones.
However a significant change in ionization potential $I$ and electron
affinity $A$ still accompanies a magic number.  Figure
\ref{fig:electronproperties}a shows the difference 
$-(I + A) / 2\approx \eps_F$.  The
value increases sharply at each magic number.

It is easy to understand that gaps at the Fermi level are associated
with an increase in stability.  The gap is created so that occupied
states are pushed down in energy while pushing unoccupied ones up,
resulting in a decrease of band structure energy.

For larger clusters that are not close to magic numbers (e.g.\
$N\approx 45$) the gap becomes small, but a significant depletion of
states around the Fermi level persists (a similar depletion of states
close to the Fermi level is also seen for the EMT-optimized clusters
e.g.\ for $N \approx 45$ and 70.  This is a product of the local
structure optimization with DFT after the EMT-annealing).  The
combined structural and electronic trends of the DFT-optimized
clusters thus point to a picture where clusters far from magic numbers
will deform significantly, maximizing the gap at the Fermi level.  In
a sense this deformation creates a new magic number for every size of
cluster provided the clusters are small enough.  As long as such a gap
remains, strong even--odd oscillations of the electronic properties
will persist due to the singly-filled state in uneven clusters.
Figure \ref{fig:electronproperties}b shows the band gap calculated as
$(I - A)/2$ as a function of cluster size.  Even and odd clusters are
plotted as separate lines.  The structure optimization tends to obtain
larger gaps close to the spherical shell closings,
and so the even--odd alternations are larger close to these. The
even--odd alternations become small compared to the 0.1\,eV smearing
for clusters larger than $\approx 40$ atoms except at the electronic
shell closings.

\begin{figure}
  \centering
  \includegraphics{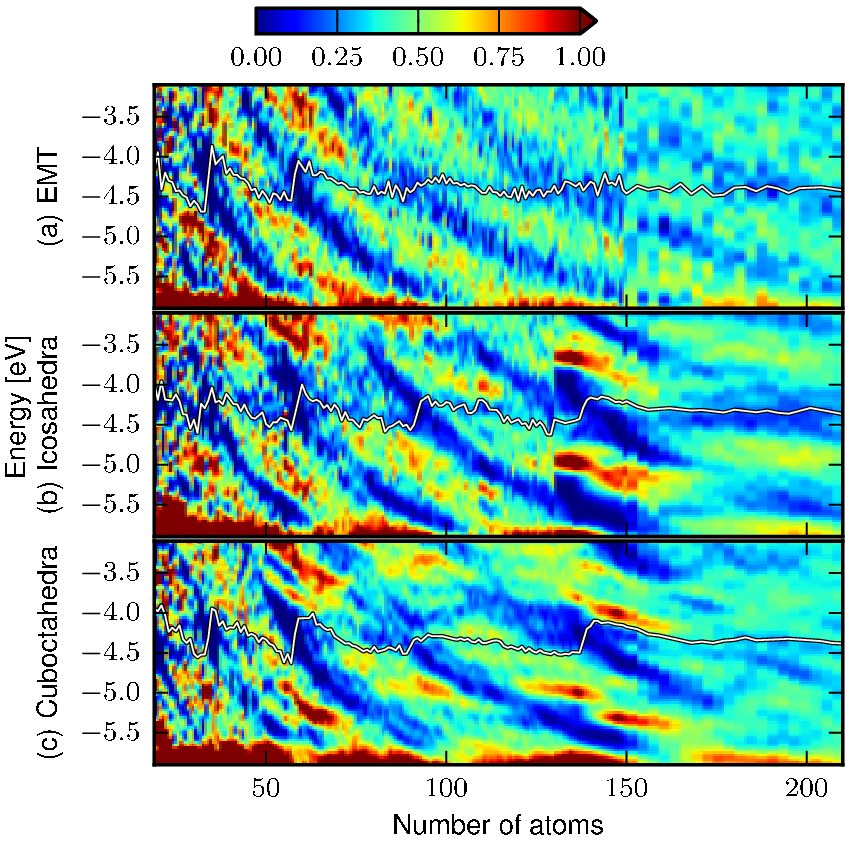}
  \colorcaption{DOS near Fermilevel for different types of clusters.  (a)
    Clusters optimized using the EMT potential.  (b) Clusters based on
    icosahedra.  (c) Clusters based on cuboctahedra.  Units are arbitrary.}
  \label{fig:manydos}
\end{figure}

Figure \ref{fig:manydos} compares the calculated DOS near the Fermi
level for EMT-optimized, icosahedral and cuboctahedral structures.
The structures yield remarkably similar electronic shells separated by
gaps.

Similarities between the electronic shell structures of spherical and
faceted structures with hundreds of electrons have previously been
found in the context of well potentials.\cite{PhysRevB.46.12649}.  Our
results show that the inclusion of a d-band, as well as the inclusion
of an atomic lattice with various irregularies as per the different
cluster generation procedures used here have limited effect on the shell
structure in this size range.

The highly visible change for icosahedra close to $N$=130 happens when
a sufficiently large number of atoms have been removed from the same
side of the cluster, causing a substantial collective movement of the
surrounding atoms (this is therefore just an artifact of the cluster
generation method).
\section{Chemical properties of clusters}

In this section we consider the chemisorption of various atoms on
cuboctahedral clusters.

Adsorption energies are calculated as follows.  A series of
cuboctahedral clusters is generated by the procedure described in
Section \ref{sec:geometry-layer-based}, so as to preserve the local
geometry around the desired adsorption site.  A structure optimization
is then performed on the entire cluster plus adsorbate, yielding a
total energy of the combined system.  From this energy we subtract the
energies of the isolated adsorbate and the isolated cluster.  Ideally
the energy of the isolated cluster should be calculated by removing
the adsorbate, then re-relaxing the structure.  However this is likely
to cause atoms to move significantly for at least some of the smaller
clusters, obscuring the general trends that we are trying to examine
without providing any insight.  For this reason we instead neglect to
re-relax the clusters after removing the adsorbate.  Calculated
binding energies therefore tend to be overestimated.

\begin{figure}
  \centering
  \includegraphics{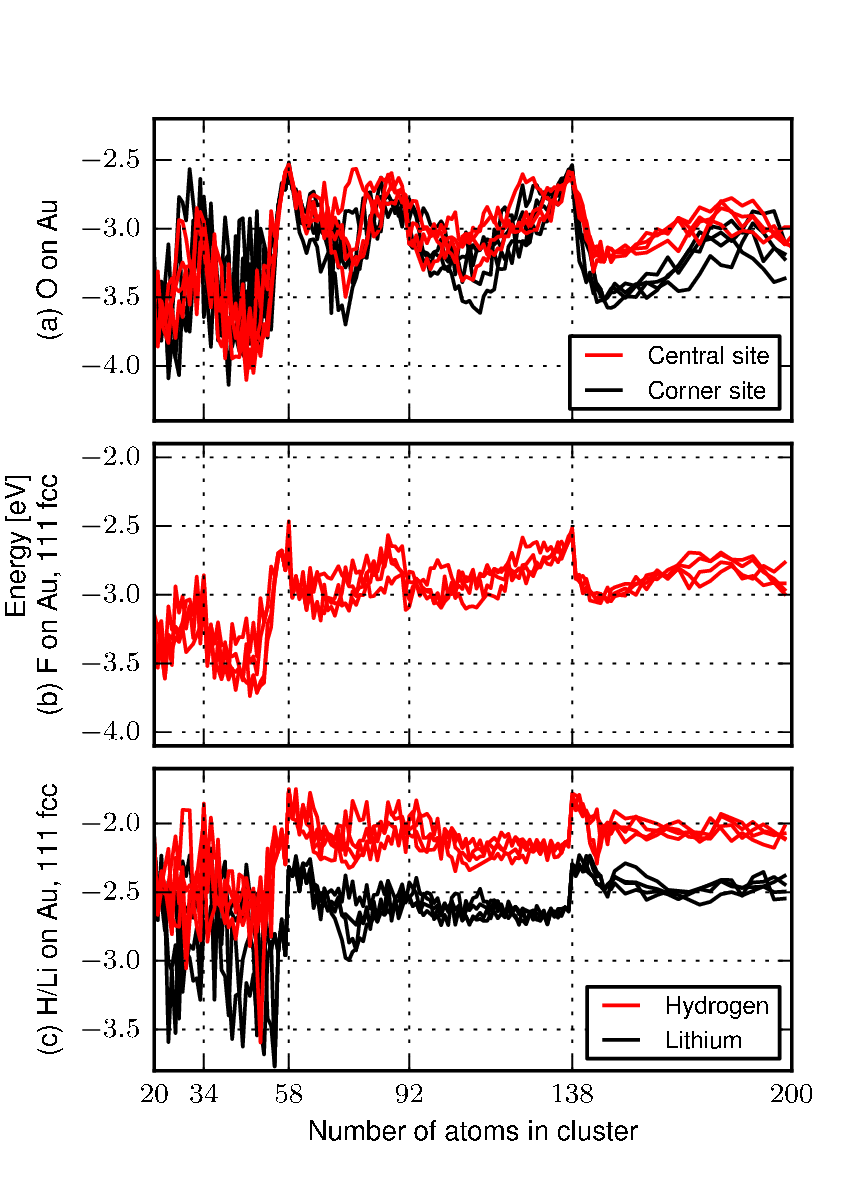}
  \colorcaption{Adsorption energies as a function of Au cluster size.  (a)
    O on the fcc site closest to the center of an (111) facet and the
    hcp site at the corner of an (111) facet.  (b) F on the central
    (111) fcc site.  (c) H and Li}
  \label{fig:adsorptionenergies}
\end{figure}

Figure \ref{fig:adsorptionenergies}a shows the adsorption energy of
oxygen on cuboctahedral clusters as a function of cluster size for two
different adsorption sites.  One is the fcc site as close as possible
to the center of an (111) facet, which locally resembles a (111) surface.
The other is the hcp site closest to the corner of a (111) facet, where
O frequently binds more strongly.  For each site there are four series
of datapoints corresponding to different random seeds in the cluster
generation procedure.

Adsorption energies on both sites are related to the distribution of
the electronic shell closings $N$=34, 58, 92 and 138 where binding
energies are particularly low.
Near the geometric shell closings 55 and 147, where the clusters are
regular and closer to being spherical, this behaviour is most pronounced.
The change near 92 is less abrupt, and we attribute this to the less 
symmetric structures generated far from geometric shell closings (we believe 
that 92 would stand out more clearly for clusters with more realistic
geometry).

Consider the behavior at the magic number 138.  The binding
gradually weakens until 138, after which it abruptly changes from very weak
to very strong.  The
same effect is seen to a smaller extent at 58 (the preceding weakening
of binding energy is in this case not gradual, but coincides with the
completion of a local facet as discussed below).

Clusters slightly larger than a magic number 
will have one or more
loosely bound electrons which can easily be donated to O.
Clusters can in this sense be characterized as alkali-like,
noble or halogen-like depending on their location relative to magic
numbers.  What the comparison between the two adsorption sites shows is that
the main variation of adsorption energy due to the electronic shell
structure is not strongly affected by local geometry around the
adsorbate.  While there are intriguing differences between the
binding on the central site and the corner site, such details are
probably too specific to make predictions about more realistic
geometries.

The alkali-like or halogen-like behavior of clusters near magic
numbers is demonstrated on Figures \ref{fig:adsorptionenergies}b and
\ref{fig:adsorptionenergies}c which show the adsorption energy of F, H
and Li on the central (111) fcc site as a function of cluster size.

The variation of F adsorption energy (Figure
\ref{fig:adsorptionenergies}b) around magic numbers is qualitatively
similar to that of O.  Since F is more electronegative, the increase
in binding energy past a magic number is more abrupt and is clearly
visible for all the magic numbers 34, 58, 92 and 138.  However F can
accept only one electron, and so the total magnitude of the increase
in energy near $N$=58 and 138 (0.5\,eV) is smaller than in the case of
O (up to 1.0\,eV).  The variation of F binding energy at magic numbers
is roughly equal to the increase in Fermi energy of the cluster
(0.6\,eV).

The electropositive Li shows the opposite trend: a steep decrease in
binding energy follows a magic number.  Again, the change in binding
energy is close to 0.5\,eV corresponding to the sharing of one
electron.  Hydrogen somewhat surprisingly follows the same trend as Li
even though the H 1s-state is approximately as low-lying as the O
2p-states.  We shall analyze this further in the next section.

Apart from the variation due to magic numbers, binding energies of all
species tend to be stronger for the smallest clusters ($N < 50$).
The very steep change in binding energy around $N$=50 
which is seen for all adsorbates can be
attributed to geometric changes of the local facet.  The triangular
6-atom (111) facet of the 55-atom cuboctahedron appears to be
generally unreactive, as all four adsorbates bind weakly to it, including O
on both the central site and the corner site.
Several previous works have noted that a lower overall coordination of nearby 
Au atoms has been found to increase binding
strengths.\cite{lopez_catalytic_2002_1,falsig_trends_2008_1,kleis-quantumsize}
The effects due to local geometry are however
comparatively small for Au clusters larger than 55 atoms.  
The global electronic shell
structure is responsible for most of the variation in adsorption
energy, as evidenced by the consistent oscillating trend.

\begin{figure}
  \centering
  \includegraphics{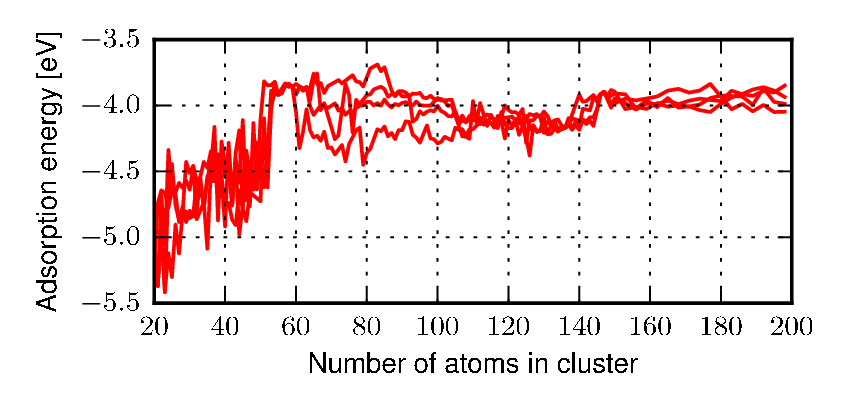}
  \colorcaption{Adsorption energy of O on Pt as a function of cluster
    size.}
  \label{fig:Pt-adsorption-energy}
\end{figure}

For comparison, adsorption energies of O on Pt clusters are shown on
Figure \ref{fig:Pt-adsorption-energy}.  The clusters have the same
initial structures as the Au clusters in Figure
\ref{fig:adsorptionenergies}a.  Instead of a smooth oscillating trend, the
binding energy can vary significantly when nearby facets are modified,
even though the modification takes place several sites away.  This
causes the adsorption energy to depend much more sensitively on the
geometry of the nearby facets.  Apart from the strongly
geometry-dependent variations, an overall decrease in O binding energy
with size is also observed, which resembles that of Au.
We note that
the changes in adsorption energy on Au clusters occur even when atoms 
are added on the directly opposite side of the cluster as seen from the 
adsorption site.

\begin{figure}
  \includegraphics{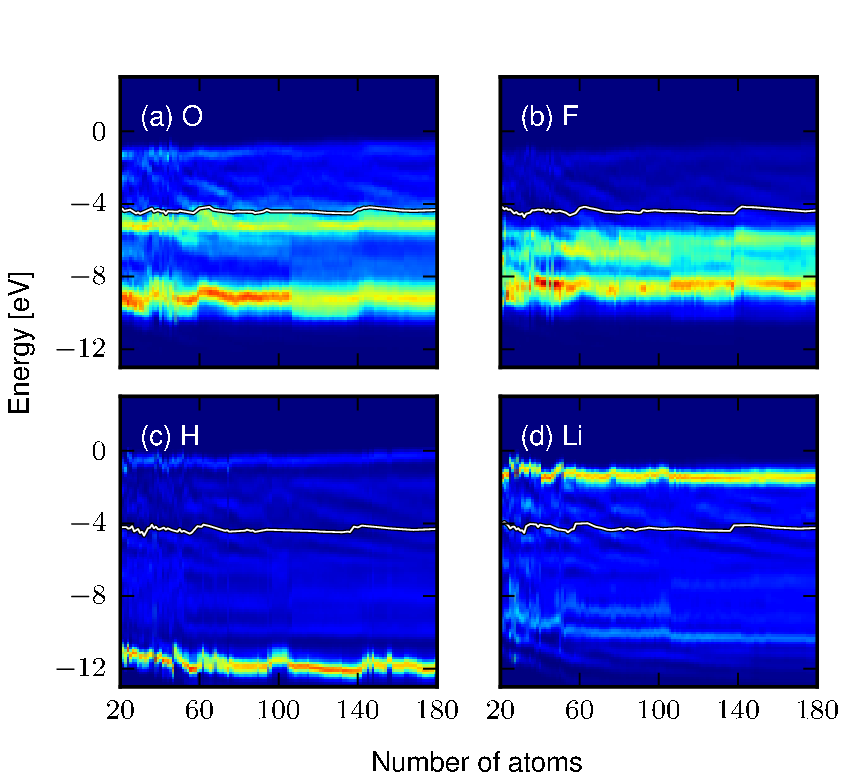}
  \colorcaption{Projected density of states on O, F, H and Li adsorbed on
    Au clusters as a function of cluster size and energy calculated
    using the atomic basis set.  The white line indicates the
    Fermi level.  Units are arbitrary.}
  \label{fig:pdos}
\end{figure}
Let us return to the Au clusters.  The projected density of states
(PDOS) on the adsorbate reveals useful information on how the atomic
states hybridize with the clusters.
The PDOS as a function of energy $\eps$ on atom $A$ is calculated from
the atomic expansion of the Kohn-Sham eigenstates $\ket{\tilde\psi_n}$
and eigenvalues $\eps_n$ as
\begin{align}
  \rho_A(\eps) = \sum_n \braket{\tilde \psi_n | P_A | \tilde \psi_n} 
  \delta(\eps - \eps_n),
\end{align}
where
\begin{align}
  P_A = \sum_{a\in A, b \in A}\ket{\Phi_a} (S_A^{-1})_{ab}\bra{\Phi_b}
\end{align}
is a projection operator onto the subspace spanned by the
basis functions $\ket{\Phi_a}$ on atom $A$, 
and $(S_A)_{ab}=\braket{\Phi_a|\Phi_b}$ 
is the overlap matrix within that subspace.

Figure \ref{fig:pdos} shows the PDOS on the O, F, H and Li atoms
adsorbed on Au cuboctahedra of varying size.  For O the p-states split
into bonding and antibonding states at the top and bottom of the
d-band (see also Ref.\ \onlinecite{hammer_why_1995_1} on hybridization
on Au surfaces).  Both bonding and antibonding states are occupied,
and they remain qualitatively similar for all sizes of clusters
although some variation is seen around the magic numbers.  The weights
of bonding and antibonding states can change drastically when the
local geometry is modified, but for clusters larger than 55 atoms these 
changes are not reflected
strongly in the adsorption energy and do not explain the trends (for
example the most visible change, around $N=105$, occurs when the last two
second-nearest neighbor atoms are added.  This causes the adsorption energy
to change by 0.15\,eV).  The behavior of F resembles that of O except
the coupling is weaker.  For H a very low-lying state is created at the
bottom of the s-band.  The surprising formation of such a low-lying state at 
the bottom of the band has been described previously for H adsorption on
Mg surfaces.\cite{1981PhRvL..46..257N}  Finally Li has a high-lying
state which is above the Fermi level.

\section{Newns-Anderson model}
The peculiar properties of metal nanoparticles are sometimes attributed
to the discrete spectrum causing the nanoparticle to behave like a
molecule rather than a bulk material characterized by a continuous
spectrum.  However the DOS quickly ($N >$ a few dozen atoms) becomes
effectively continuous on any reasonable energy scale ($\sim 0.1$\,eV).  
The primary feature distinguishing clusters with a few
dozens to a few hundred atoms compared to bulk is not whether the DOS
is discrete or continuous, but rather the fact that the approximately
continuous DOS remains grouped into shells separated by gaps.  The
size-dependent variation in this effectively continuous DOS, and in
particular the distribution of magic numbers, are the significant
factors that make clusters with many hundreds of atoms still differ
from bulk Au.  Since the DOS
of an Au cluster larger than a few dozen atoms can effectively be regarded as
continuous, we will in the following apply the Newns-Anderson 
model\cite{Newns-PhysRev.178.1123} to
understand chemisorption on Au clusters.  The Newns-Anderson model considers 
a single state $\ket
a$ on an atom which hybridizes with a continuum of states $\{\ket k\}$
of the metal surface described by a Hamiltonian of the form
\begin{align}
  H = H_0 + V,
\end{align}
where $H_0$ is the Hamiltonian of the uncoupled cluster plus
adsorbate, and $V$ describes the coupling.  In the basis consisting of
metal eigenstates $\{\ket k\}$ plus adsorbate eigenstate $\ket a$, the
uncoupled Hamiltonian $H_0$ is diagonal, and the Newns-Anderson
Hamiltonian can be written in block form as
\begin{align}
  H = \left[
  \begin{array}{ccc|c}
    \ddots &        & 0       & \vdots \\
           & \eps_k &         & v_{ka}^*\\
    0      &        & \ddots  & \vdots \\
    \hline
    \cdots & v_{ak} & \cdots   &\eps_a
  \end{array}
  \right].\label{eq:newnsandersonhamiltonian}
\end{align}
Here $\eps_a$ and $\eps_k$ are the uncoupled energy eigenvalues on
the atom and in the metal while $v_{ak}=\braket{a|V|k}$ are coupling
matrix elements.

Our idea now is to perform a DFT calculation to obtain a Hamiltonian
matrix of the composite system consisting of both cluster and
adsorbate, then transform it to Newns-Anderson form cf.\ Eq.\
\eqref{eq:newnsandersonhamiltonian}.  In the basis of atomic orbitals
used by GPAW, the Kohn-Sham equations are solved as a generalized eigenvalue
problem\cite{larsen_localized_2009}
\begin{align}
  \sum_\nu H_{\mu\nu} c_{\nu n} = \sum_\nu S_{\mu\nu} c_{\nu n} \eps_n.
\end{align}
The overlap matrix $S_{\mu\nu}$ is present because the basis is
non-orthogonal.  The Hamiltonian can be regarded as a blocked matrix
\begin{align}
  H^{\mathrm{DFT}} = 
  \begin{bmatrix}
    H^M     & H^{MA}\\
    H^{AM}  & H^A
  \end{bmatrix}
\end{align}
with one block $H^M$ corresponding to the metal or cluster, one block
$H^A$ corresponding to the atom and the off-diagonal blocks $H^{MA},
H^{AM}$ corresponding to the interaction.  We diagonalize each of the
metallic and atomic submatrices according to
\begin{align}
  \sum_{\nu} H_{k'\nu}^M c_{\nu k}^M 
  &= \sum_{\nu} S_{k'\nu}^M c_{\nu k}^M \eps_k,\\
  \sum_{\nu} H_{a'\nu}^A c_{\nu a}^A 
  &= \sum_{\nu} S_{a'\nu}^A c_{\nu a}^A \eps_a,
\end{align}
to obtain values for the energies $\eps_k$ and $\eps_a$ of the
uncoupled systems.  Clearly these are approximate, as the real energy
values could be evaluated selfconsistently on each of the uncoupled
systems using a separate DFT calculation.  However the Hamiltonian and 
eigenvalues from one selfconsistent calculation cannot be expected to be 
``compatible'' with
those from a different selfconsistent calculation.  Indeed it is known
from the force 
theorem\cite{mackintosh1980electrons,PhysRevB.35.7423}
that the first-order change in energy due to a small 
perturbation
of the Hamiltonian is equal to the change in band structure energy,
keeping the potential and density fixed.  Different Hamiltonian
matrices based on DFT can therefore be expected to contain all
information provided that they are constructed from the same potential
and density.

Using the matrices $c^M$ and $c^A$, coupling elements can be obtained
through the transformation
\begin{align}
  v_{ak} &= \sum_{a' k'} c_{a'a}^{A*} H_{a'k'}^{AM} c_{k'k}^M,\\
  s_{ak} &= \sum_{a' k'} c_{a'a}^{A*} S_{a'k'}^{AM} c_{k'k}^M.
\end{align}
Thereby we have all the parameters in the Newns-Anderson Hamiltonian
(Eq.\ \eqref{eq:newnsandersonhamiltonian}), although the adsorbate
state has an overlap $s_{ak}=\braket{a|k}$ with each of the metallic
eigenstates, meaning the basis set is non-orthogonal.  Below we will
use expressions derived by Grimley\cite{grimley_overlap_1970} for
the non-orthogonal case.

If the adsorption induces a change $\delta \rho(\eps)$ in the metallic
density of states, the adsorption energy can be written as
\begin{align}
  E_{\mathrm{ads}} = 2 \int_{-\infty}^{\eps_F}
  \delta\rho(\eps)\eps\idee\eps - \Delta N \eps_F + n_a (\eps_F - \eps_a),
  \label{eq:deltarho-adsorption-energy}
\end{align}
where $\eps_F$ is the Fermi level.  The first term is the contribution
to the binding energy from hybridization with the adsorbate (the
factor of 2 assumes that each spin hybridizes equally and independently).  The
integral of the induced density of states
\begin{align}
  \Delta N = 2 \int_{-\infty}^{\eps_F} \delta \rho(\eps)\idee\eps
\end{align}
is the change in number of electrons below the Fermi level.  If this
integral is nonzero, either too much or too little charge will be
counted in the integration up to the Fermi level, and the extra or
missing electrons must then be deposited onto or taken from the Fermi
level.  This electron count correction is the second term, $\Delta N
\eps_F$.  Finally if the atom contributes $n_a$ electrons which come
from the adsorbate level $\eps_a$, this amount of extra charge must in
turn be deposited on the Fermi level $\eps_F$ (last term).  In the
Newns-Anderson model, the first two terms of Eq.\
\eqref{eq:deltarho-adsorption-energy} are expressed as the integral
over a function $\eta(\eps)$ such that
\begin{align}
  E_{\mathrm{ads}} = \frac{2}{\pi} \int_{-\infty}^{\eps_F} \eta(\eps) \idee \eps 
  + n_a (\eps_F - \eps_a),\label{eq:na-adsorption-energy}
\end{align}
where
\begin{align}
  \tan \eta(\eps) &= \frac{\Delta(\eps, \eps)}{\eps - \eps_a - \Lambda(\eps)},\\
  \Lambda(\eps) &= \frac 1 \pi \int_{-\infty}^\infty
  \frac{\Delta(\eps, \eps')}{\eps - \eps'} \idee \eps',\\
    \Delta(\eps, \eps') 
  &= \pi \sum_k |\eps s_{ak} - v_{ak}|^2 \delta(\eps' - \eps_k).
\end{align}
The one-variable function $\Delta(\eps) \equiv \Delta(\eps, \eps)$ is
referred to as the \emph{chemisorption function} and plays the role of
a continuous coupling matrix element.
$\eta(\eps)$ is the phase shift of the complex self-energy
$\Lambda(\eps) - i\Delta(\eps)$ and is related to the the
induced density of states by
\begin{align}
  -\frac 1\pi \diff{\eta(\eps)}{\eps} = \delta\rho(\eps).
\end{align}

\begin{figure}
  \centering
  \includegraphics{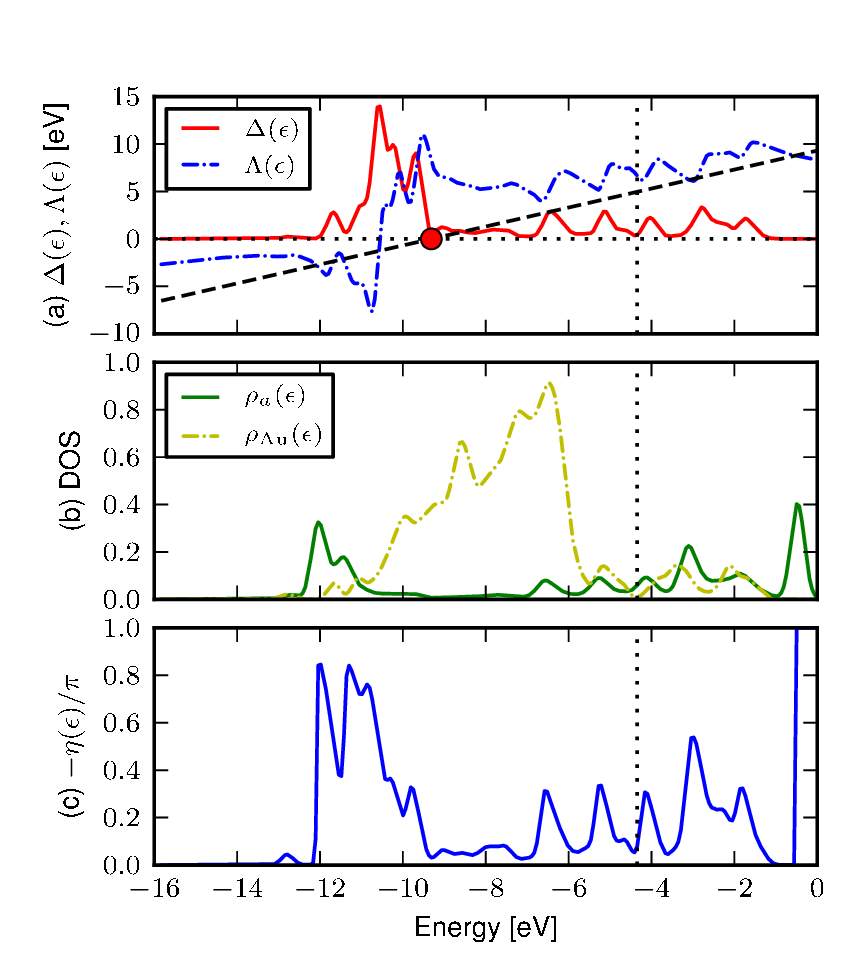}
  \colorcaption{(a) $\Delta(\eps)$, $\Lambda(\eps)$ and $\eps-\eps_a$ for H
    on 58-atom Au cluster.  The dotted line indicates the Fermi level.  
    (b) Projected DOS on atom in eV$^{-1}$ and total DOS of 
    isolated cluster in arbitrary units. (c) Cumulative induced DOS.}
  \label{fig:na-overlap-H}
\end{figure}
The PDOS on the adsorbate can be written as
\begin{align}
  \rho_a(\eps) = \frac 1 \pi
  \frac{\Delta(\eps)}{(\eps - \eps_a - \Lambda(\eps))^2 + \Delta^2(\epsilon)}.
  \label{eq:na-pdos}
\end{align}

Because of the approximations used in this method, calculated binding
energies are by themselves not useful (or accurate) compared to the
DFT results.  The strength of this method lies in the conceptual
simplification that the binding energy can be understood from
continuous functions such as $\Delta(\eps)$ and $\eta(\eps)$.  This allows
the origin of the coupling and binding energy to be attributed
to particular states in the cluster.  Next we will apply this to H, O,
Li and F on a 58-atom Au cluster to understand the effect of magic
numbers on chemisorption.

\subsection{Adsorption of H}
We perform a DFT calculation on a 58-atom
Au cluster with H adsorbed to obtain the Hamiltonian and overlap
matrix.  In this calculation, only the 1s basis function is included
on H, but otherwise the parameters are identical to those used in earlier
calculations.  For the 1s basis function we calculate $\Delta(\eps)$
and $\Lambda(\eps)$ which is shown on Figure \ref{fig:na-overlap-H}a
together with the line $\eps-\eps_a$.  When $\eps-\eps_a =
\Lambda(\eps)$ and $\Delta(\eps)$ is small, there will be resonances
in $\rho_a(\eps)$ (as per Eq.\ \eqref{eq:na-pdos}) corresponding to states
on the atom.  The adsorbate level $\eps_a$
(circle) and Fermi level (dotted line) are indicated.
The resulting PDOS is shown
on Figure \ref{fig:na-overlap-H}b together with the total DOS of the
Au cluster.  This reveals that it is the strong coupling to low-lying
metallic states ($\eps\approx-11$\,eV) which gives rise to a bonding,
localized state at the bottom of the Au s-band, at $-12$\,eV, and an
antibonding state consisting of several peaks mostly above the Fermi
level.

\begin{figure}
  \centering
  \includegraphics{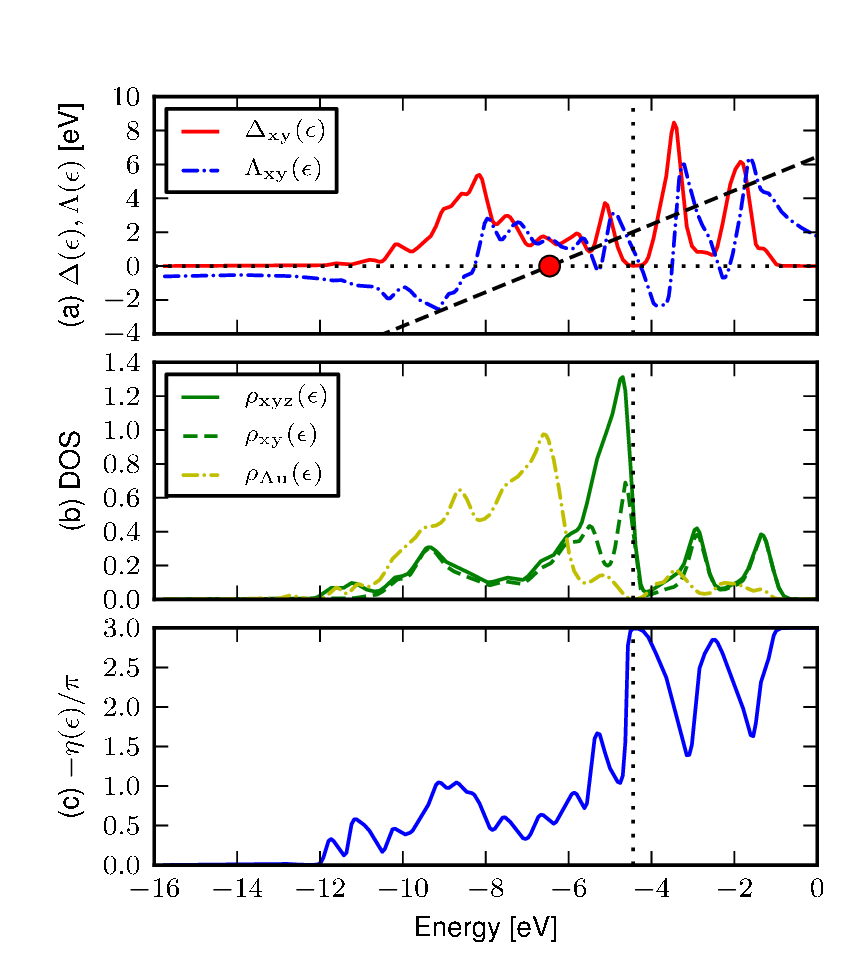}
  \colorcaption{(a) $\Delta(\eps)$, $\Lambda(\eps)$ and $\eps - \eps_a$ for
    O p$_x$ or p$_y$ on 58-atom Au cluster.  
    $\eps_a$ is indicated on the energy axis.
    (b) Projected DOS on all O 2p states in eV$^{-1}$
    (solid line), on p$_x$ and p$_y$ only (dashed line), and total DOS
    of isolated cluster in arbitrary units.  (c) Cumulative induced DOS.}
  \label{fig:na-overlap-O}
\end{figure}

On a side note, the very high PDOS at the antibonding state may seem
surprising.  Given that the adsorbate level $\eps_a\approx-9$ is much
closer to the bonding state, the bonding state would be expected to be
similar to $\ket a$, and thus have a high PDOS on the atom, while
the antibonding state should be more similar to the metallic states
and therefore have a low PDOS on the atom.  However the inclusion of
overlap $s_{ak}$ in the model causes part of the states on the
neighboring metal atoms to be counted in the adsorbate PDOS,
contributing to the prominence of the antibonding peak.  While the
overlap affects the calculated PDOS, the overlap is correctly
taken into account in binding energies and other parts of the
formalism.

In the creation of bonding and antibonding states the original
adsorbate state has been eliminated, and a change
$\delta\rho(\eps)$ in DOS has been induced in the cluster.  The
cumulative induced DOS $-\eta(\eps)/\pi$ is shown on Figure
\ref{fig:na-overlap-H}c.  While the newly created bonding state at
$-12$\,eV can accept a certain amount of charge, a similar amount of
charge has been removed from the remainder of the cluster DOS (mostly
around $-10$\,eV) such that the total integral of the induced DOS up to
the Fermi level is zero.  The extra electron from the H atom is
therefore deposited on the Fermi level.  A higher-lying Fermi level
implies a weaker adsorption energy, since the electron is deposited at
a higher energy.  This is why clusters just past a magic number,
characterized by a higher Fermi level but an otherwise similar
spectrum, adsorb H more weakly than clusters just before a magic
number.  We can also see how the induced DOS integrates to zero only
because the Fermi level is located at a gap between electronic shells:
Within each electronic subshell there are fluctuations in the induced
DOS which correspond to slight movements of the electronic shells but
without the introduction of any extra charge.  These cause
$\eta(\eps)$ to locally deviate from 0.  The adsorption energy
therefore may not depend simply on the Fermi energy in general, but
must do so at the magic numbers.
These results are consistent with previous findings for very small 
clusters, that H atoms effectively contribute
their electron to the LUMO, behaving
like an extra Au atom.\cite{buckart_anomalous_2003,Phala200433}

\subsection{Adsorption of O}
Since the Newns-Anderson model only takes a single
state into account while O has three p-states, we will assume that
each of the states hybridizes independently and contributes to the
adsorption energy as per Eq.\ \eqref{eq:na-adsorption-energy}.  Thus
we consider one Hamiltonian of the form
\eqref{eq:newnsandersonhamiltonian} for each p-state with varying
$v_{ak}$ and $s_{ak}$.

The 2p$_x$ and 2p$_y$-states
are close to degenerate and
have almost identical chemisorption functions.  Figure \ref{fig:na-overlap-O}a shows the average $\Delta(\eps)$ and
$\Lambda(\eps)$ from the O 2p$_x$ and 2p$_y$-states.
In this case the
weaker splitting leads to greater smearing of the states close to the
d-band and between the electronic shells.  The higher-lying peaks in 
$\Delta(\eps)$ correspond to coupling with of the electronic shells.  Figure
\ref{fig:na-overlap-O}b shows the total PDOS $\rho_{xyz}(\eps)$ due
to all three p-states (full lines) along with the contribution
$\rho_{xy}(\eps)$ from p$_x$ and p$_y$ (dashed).  The most profound
feature is the state between the top of the d-band and the Fermi
level, which therefore is filled.  The induced DOS integrates
to 3.0 at the Fermi level (Figure \ref{fig:na-overlap-O}c),
allowing space for six electrons counting spin-degeneracy.  Since
only four electrons are contributed, a total of two electrons are
taken from the Fermi level into available lower-lying states.  An
upward shift in Fermi level therefore implies that more energy is
gained from this transfer, causing a change in binding energy
opposite that for H adsorption as seen from the DFT calculations.

\subsection{Adsorption of Li and F}
\begin{figure}
  \centering
  \includegraphics{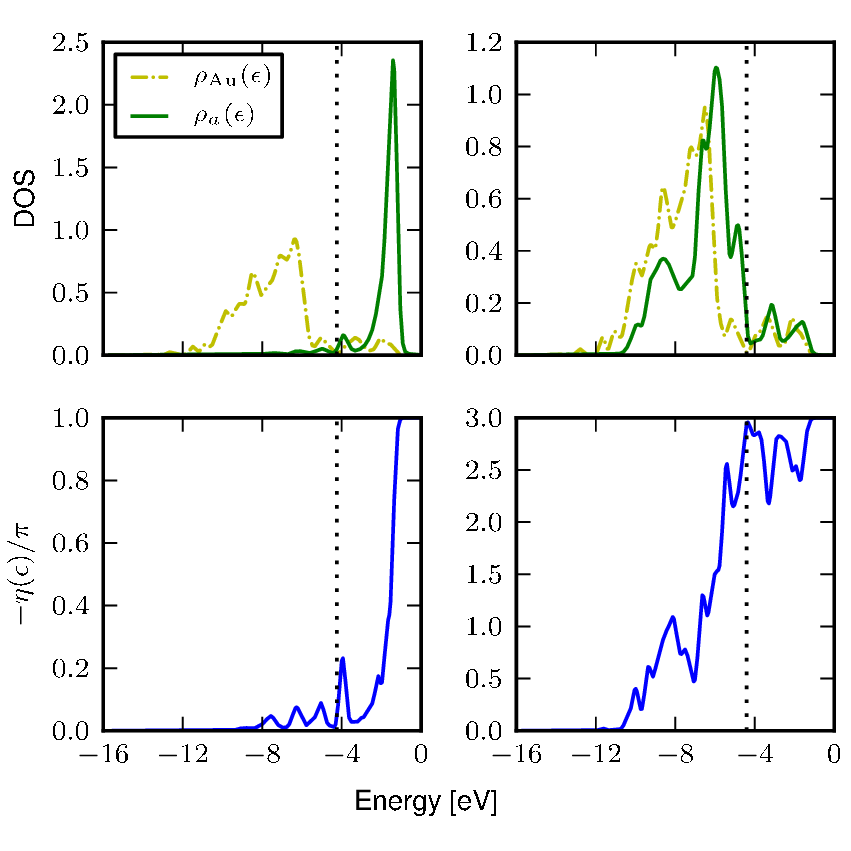}
  \colorcaption{Projected density of states (top) on adsorbate (eV$^{-1}$)
    and cluster (arbitrary units), and
    cumulative induced density of states (bottom) for Li (left) and F
    (right).}
  \label{fig:na-overlap-multiplespecies}
\end{figure}

Li is the simplest of the four cases.  Here the bonding is weak enough
that no significant splitting occurs.  The adsorbate state instead
broadens into a resonance far above the Fermi level, see the left part
of Figure \ref{fig:na-overlap-multiplespecies}, without inducing any
states below the Fermi level.  The electron contributed by the Li atom
therefore moves down to the Fermi level causing the same dependence of
adsorption energy on Fermi level as for H.

F couples more weakly than O, and the bonding states are therefore split up
less than for O (right part of Figure \ref{fig:na-overlap-multiplespecies}).
Since both bonding and antibonding states are occupied, F
behaves like O except only one electron can be transferred from the
Fermi level, meaning that the change in adsorption energy at
magic-number clusters is generally smaller than for O.

\subsection{Comparison to Pt clusters}
\begin{figure}
  \includegraphics{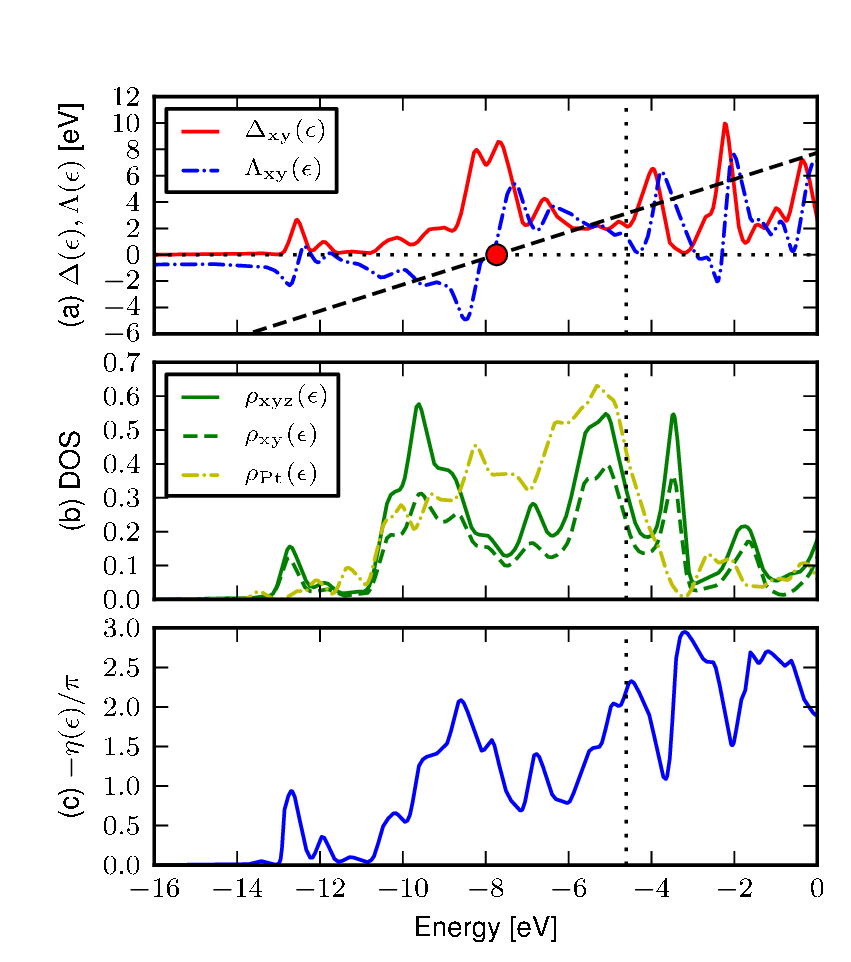}
  \colorcaption{Chemisorption of O on 58-atom Pt cluster
    in comparison with Au from Figure \ref{fig:na-overlap-O}.}
  \label{fig:na-Pt}
  \centering
\end{figure}

Finally we shall briefly consider the binding of O to the 58-atom Pt
cluster.  Again the p$_x$ and p$_y$ states are close to degenerate,
and the average of their chemisorption functions is shown on Figure
\ref{fig:na-Pt}a.  Two primary features appear in the chemisorption
function: a strong coupling within the d-band around $\eps=-8$\,eV,
and a number of higher-lying peaks corresponding to electronic shells
like those of Au clusters.  Due to the broader and higher-lying
d-band, the adsorbate state splits into peaks over a wider energy
range as seen on Figure \ref{fig:na-Pt}b.  The increase in binding on
Pt compared to Au manifests itself as an increase in area below the
curve in Figure \ref{fig:na-overlap-O}c cf.\ Eq.\
\eqref{eq:na-adsorption-energy}.  The overall upward shift of the
coupling leads to an upward shift of the induced density of states,
and so only approximately 2.2 out of the 3 states contributed by O are
in this case located below the Fermi level.  The partial occupation of O 
2p-states has been studied and confirmed
experimentally.\cite{strasser_lattice-strain_2010_1}
Recall that for Au, essentially all of the adsorbate-induced states
were located below the Fermi level (Figure \ref{fig:na-overlap-O}c).

The s-electron shell structure is mostly visible in the chemisorption
function well above the Fermi level.  Further, since the location of
the Fermi level within the d-band prevents abrupt changes in the Fermi
level with cluster size, the s-electron shell structure---as expected
---cannot exert a strong influence on the chemical binding on Pt
clusters.

\section{Conclusions}
The structure of very small Au clusters is intricately dependent on
the s-electron hybridization, with clusters at magic numbers having
very large band gaps.  Clusters with different numbers of atoms deform
considerably to minimize the band structure energy by creating gaps at
the Fermi level.  For odd-numbered clusters this results in a
singly-filled state in the middle of the gap, causing strong even--odd
oscillations of HOMO and LUMO.

Clusters based on regular geometries or a simple EMT potential show a
more clear electronic shell structure and have large band gaps only at
the major shell closings.  These structures are less realistic, but
are computationally feasible to optimize for larger cluster sizes.

Adsorption energies of atoms on regular Au clusters oscillate with the
electronic magic numbers.  While local geometry is
known to be important, the variation in binding energy of O due to
magic numbers alone may be up to 1 eV.  Clusters just before or after
magic numbers are found to exhibit roughly halogen-like and
alkali-like behavior while magic-number clusters are, as expected,
universally unreactive.

A more detailed analysis attributes the increase or decrease in
binding energy of specific adsorbates at magic numbers to properties of the
adsorbate-induced density of states.  Adsorption of O or F induces
states below the Fermi level, allowing the transfer of electrons from the
Fermi level into the lower-lying states.  In contrast H and Li, despite
having very different adsorbate levels and electronegativity, only
induce states above the Fermi level, and the electron contributed by
these atoms is therefore transferred to the Fermi level.

Center for Atomic-scale Materials Design is funded by the Lundbeck
Foundation.  The Catalysis for Sustainable Energy initiative is funded
by the Danish Ministry of Science, Technology and Innovation.  We
acknowledge support from the Danish Center for Scientific Computing,
from the Danish Council for Strategic Research's Programme Commission
on Strategic Growth-Technologies (NABIIT), and from the US Department
of Energy---Basic Energy Sciences through the SUNCAT Center for
Interface Science and Catalysis.

\end{document}